%% file: P5HBSeidel.tex
\documentclass{amsart}
\usepackage[T1]{fontenc}
\usepackage[latin1]{inputenc}
\usepackage{graphicx}
\usepackage{epic,eepic}
\usepackage[normalem]{ulem}
\newtheorem{thm}{Theorem}[section]
\newtheorem{cor}[thm]{Corollary}

\newtheorem{prop}[thm]{Proposition}

\newtheorem{clm}{Claim}[thm]
\newtheorem{rem}[thm]{Remark}
\newtheorem{notation}[thm]{Notation}
\theoremstyle{definition}
\newtheorem{defn}[thm]{Definition}
\theoremstyle{remark}

\numberwithin{equation}{section}
\newenvironment{prf}{{\bf \noindent Proof } }{\hfill$\square$\\}
\newenvironment{prfClaim}{{\bf Proof }}{{\hfill\tiny{$\square$\\}}}

\newcommand{\ligne}{\vspace{0,5cm}}
\newcommand{\ignore}[1]{}
\title[]{Seidel complementation on ($P_5$, $House$, $Bull$)-free graphs}
\author{J.L. Fouquet and J.M. Vanherpe}
\address{L.I.F.O., Facult\'e des Sciences, B.P. 6759 \\
Universit\'e d'Orl\'eans, 45067 Orl\'eans Cedex 2, FR}

\subjclass{035 C} \keywords{}

\date{}

\begin{document}
\input{epsf.sty}
\begin{abstract}

\end{abstract}
\maketitle

\section{Seidel complementation}
\begin{notation}\label{Not:VoisinagesEtAutres}
$N(v)$ denotes the neighborhood of the vertex $v$, $N[v]=N(v)\cup\{v\}$ and $\overline{N(v)}=V-N(v)$.
\end{notation}

Given a graph $G=(V,E)$ and a vertex $v$ of $G$, the seidel complement of $G$ on $v$ inverses all edges between $N(v)$ and $V-N[v]$. More formaly
\begin{defn}\label{def:SeidelComplement}
Let $=(V,E)$ be an undirected graph and let $v$ be a vertex of $G$. The Seidel complement oat $v$ on $G$, denoted $G*v$ is defined as follows~:\\
$G=v=(V, E_1\cup E_2\cup E_3)$ where $E_1=\{xy|xy\in E, x\in N[v], y\in N[v]\}$, $E_2=E\cap \overline{N(v)}^2$ and $E_3=\{xy|xy\notin E, x\in N(v), y\in \overline{N(v)}\}$.
\end{defn}
\begin{rem}\label{Rem:PropretiesSeidelComplementation}\cite{Lim09}
\begin{itemize}
\item $G*v*v=G$.
\item If $G$ is a cograph and $v$ is a vertex of $G$ then $G*v$ is a cograph.
\item $G$ is prime with respect to modular decomposition if and only if $G*v$ is prime with respect to modular decomposition
\item $\overline{G*v}=\overline{G}*v$
\end{itemize}
\end{rem}
\section{seidel complementation on ($P_5$, $\overline{P_5}$,$Bull$)-free graphs}
\begin{figure}
\input{P5HB-free.eepic}
\caption{Forbidden configurations for ($P_5$, $House$, $Bull$)-free graphs.}\label{fig:P5HB-free}
\end{figure}
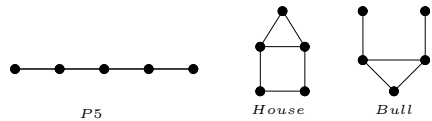
\begin{thm}\label{thm:SeidelComplementationOnP5HB-free}
A graph $G$ is ($P_5$, $House$, $Bull$)-free if and only if for all vertex $v$ of $G$, $G*v$ is ($P_5$, $House$, $Bull$)-free.
\end{thm}
\begin{prf}
Let $v$ be a vertex of $G$. Assume that $G*v$ contains an induced subgraph, say $H$,which is  isomorphic to either a $P_5$ or a $House$ or a $Bull$.
\begin{clm}\label{clm:vPasSurP5NiHouseNiBull}
The vertex $v$ does not belong to $H$.
\end{clm}
\begin{prfClaim}
Assume not. Figure \ref{fig:vSurH} describes all cases that can occur whenever $v$ is a vertex of $H$. It is not difficult to check that, in all cases, $(G*v)*v$ would contain a subgraph isomorphic to $P_5$ or to $House$ or to $Bull$, a contradiction since $G*v*v=G$ and $G$ is assumed to be ($P_5$, $House$, $Bull$)-free.

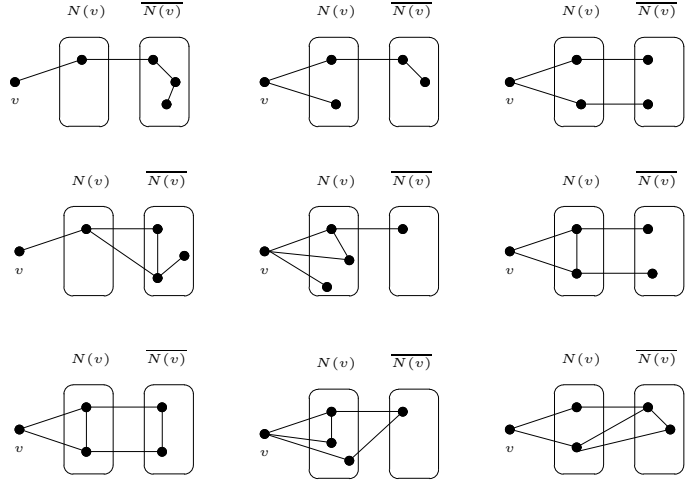
\begin{figure}
\input{SommetvSurH.eepic}
\caption{The possible cases when $v$ is a vertex of $H$.}
\label{fig:vSurH}
\end{figure}
\end{prfClaim}
In the following we suppose that the vertices of $H$ are in $N(v)\cup \overline{N(v)}$, moreover $H$ has vertices in both sets $N(v)$ and $\overline{N(v)}$, otherwise $H$ would be an induced subgraph of $G*v*v=G$, a contradiction.

Figure \ref{fig:SommetvPasSurP5} (resp Figure \ref{fig:SommetvPasSurBull}, Figure \ref{fig:SommetvPasSurHouse}) describes all cases that can occur whenever $H$ is isomorphic to a $P_5$ (resp. a $Bull$, a $House$) and has at most two vertices in $N(v)$. In all cases we get a contradiction with Claim \ref{clm:vPasSurP5NiHouseNiBull} or a forbidden configuration appears in $G*v*v$.
\begin{figure}
\input{SommetvPasSurP5.eepic}
\caption{Cases~: $H$ is a $P_5$ and has at most $2$ vertices in $N(v)$.}\label{fig:SommetvPasSurP5}
\end{figure}
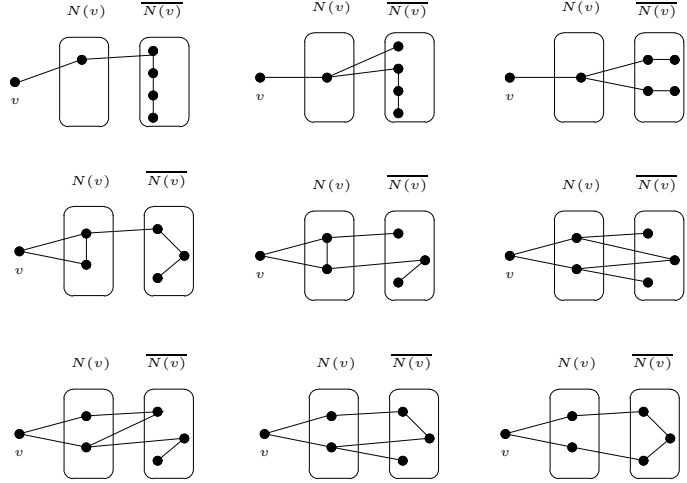
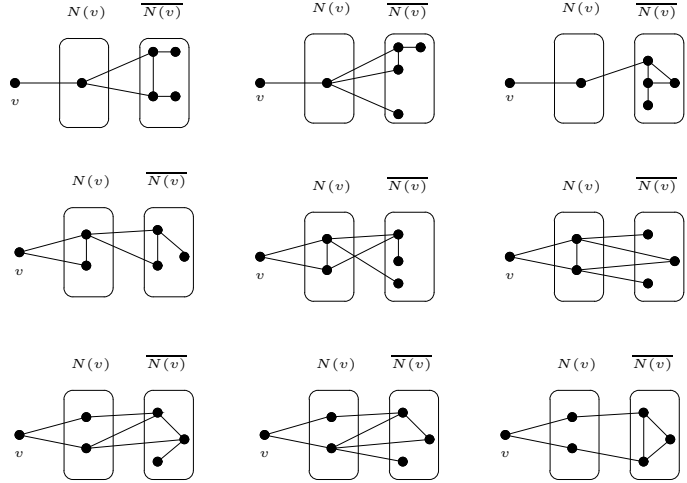
\begin{figure}
\input{SommetvPasSurBull.eepic}
\caption{Cases~: $H$ is a $Bull$ and has at most $2$ vertices in $N(v)$.}\label{fig:SommetvPasSurBull}
\end{figure}
\begin{figure}
\input{SommetvPasSurHouse.eepic}
\caption{Cases~: $H$ is a $House$ and has at most $2$ vertices in $N(v)$.}\label{fig:SommetvPasSurHouse}
\end{figure}
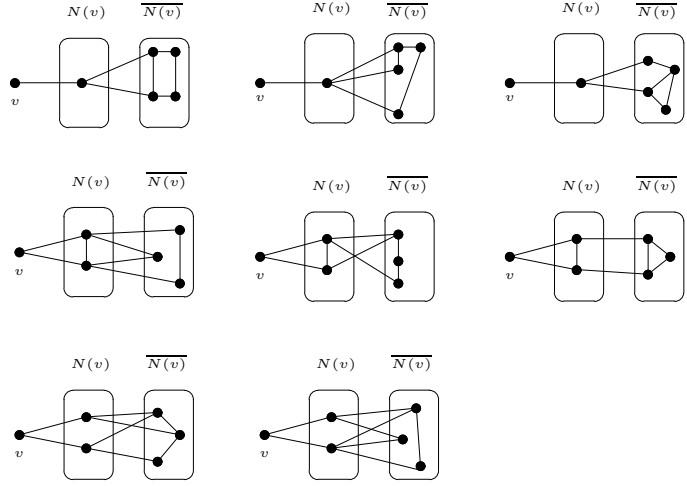
When $H$ has more than $2$ vertices in $N(v)$ we get a similar contradiction in considering $\overline{G}$.
\end{prf}
\section{seidel complementation on the modular decomposition tree of ($P_5$, $\overline{P_5}$,$Bull$)-free graphs}
\subsection{On ($P_5$, $\overline{P_5}$)-free graphs}
\ligne We shall say that an induced subgraph of a ($P_5$,$\overline(P_5)$)-free graph is a {\em buoy}\cite{Fou93} whenever we can find a partition of its vertex set into $5$ subsets $A_i,i=1,\ldots,5$ (subsript $i$ is to be taken modulo $5$, such that $A_i$ and $A_{i+1}$ are joined by every possible edge, while no possible edge are allowed between $A_i$ and $A_j$ when $j\neq i+1$ $ mod[5]$, and such that the $A_i$'s are maximal for these properties.

\begin{thm}\label{thm:TheoremeDeLaBouee}\cite{Fou93,FouGiaThuMai95}
Let $G$ be a connected ($P_5$,$\overline{P_5}$)-free graph. If $G$ contains an induced $C_5$ then every $C_5$ of $G$ is contained into a buoy, and this buoy is either equal to $G$ or is an homogeneous set of $G$.
\end{thm}
\begin{cor}\label{cor:C5ouC5Free}
Every prime ($P_5$,$\overline{P_5}$)-free graph is either a $C_5$ or is $C_5$-free
\end{cor}

\ignore{
\ligne  Efficient algorithms for ($P_5$,$\overline{P_5}$,$C_5$)-free graphs are due to Hoang in \cite{Hoang}. Those algorithms and Corollary $2.1$ are used in \cite{GiakRusu} to produce $O(n(n+m))$ time complexity algorithms for finding a maximal weighted clique or a maximal weighted stable set for ($P_5$,$\overline{P_5}$)-free graphs.

Observe that ($P_5$,$\overline{P_5}$,$C_5$)-free graphs are perfectly orderable graphs (\cite{Hoang2}). But the algorithms proposed in \cite{Chvatal} by Chv\'atal et al. cannot be used since we need, as shown below, the solutions for weighted versions on prime graphs even for the ``simple'' unweighted optimization problems.
}
\subsection{On ($P_5$, $\overline{P_5}$, $Bull$)-free graphs}
\begin{thm}\label{thm:StructureDesPrimeP5HB-free}\cite{Fou93} Let $G$ be a prime graph.
$G$ is a $P_5HB$-free graph if and only if one of the following conditions is satisfied
\begin{itemize}
\item[(i)] $G$ is isomorphic to a $C_5$;
\item[(ii)] $G$ is bipartite and $P_5$-free;
\item[(iii)] $\overline{G}$ is bipartite and $P_5$-free.
\end{itemize}
\end{thm}
In a prime $P_5$-free biparte graph  the neighborhoods of two distinct vertices cannot overlap properly, thus~:
\begin{prop}\label{prop:StructurePrimeP5FreeBiparti}
Let $G=(V,E)$  be a prime graph of $n$ vertices.
 $G$  is bipartite and $P_5$-free
iff the following conditions are verified:
\begin{itemize}
\item[(i)] There exists a partition of $V(G)$ into two stable sets $B=\{b_1,b_2,\ldots b_{\frac{n}{2}}\}$ and $W=\{w_1,w_2,\ldots w_{\frac{n}{2}}\}$.
\item[(ii)] The neighbors of $b_i$ ($i=1\ldots \frac{n}{2}$) are precisely $w_1,\ldots w_{\frac{n}{2}-i+1}$.
\end{itemize}
\end{prop}
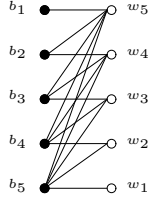
\begin{figure}
\input{P5-freeBiparti.eepic}
\caption{A Prime $P_5$-free bipartite graph.}\label{fig:PrimeP5FreeBiparti}
\end{figure}
\subsection{Seidel complementation of a Prime $P_5$-free bipartite graph}
\begin{figure}
\input{SeidelComplementationOnb2.eepic}
\caption{Seidel complementation on $b_2$.}\label{fig:SeidelComplementationPrimeP5FreeBiparti}
\end{figure}
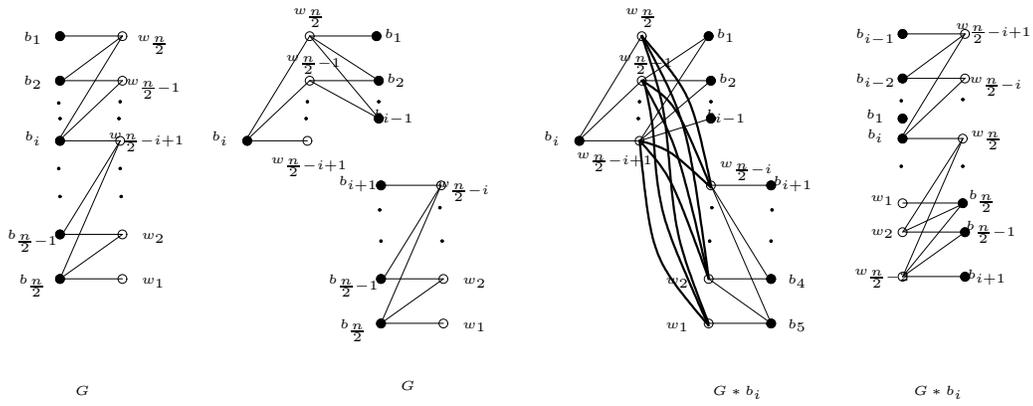
\begin{prop}\label{prop:SeidielComplemetationOnbi}
Let $G=(B\cup W,E)$ be a prime bipartite $P_5$-free graph such that $B=\{b_1,\ldots b_{\frac{n}{2}}\}$ and $W=\{w_1,\ldots,w_{\frac{n}{2}}\}$, then $G*b_i$ is a prime $P_5$-free bipartite graph together with the bipartition~: $$B'=\{b_{i-1},b_{i-2},\ldots b_i, w_1,w_2,\ldots,w_{\frac{n}{2}}\}, W'=\{b_{i+1},\ldots, b_{\frac{n}{2}}, w_{\frac{n}{2}}, w_{\frac{n}{2}-1},\ldots w_{\frac{n}{2}-i+1}\}$$
\end{prop}
\begin{prf}
We have $N(b_i)=\{w_{\frac{n}{2}}, w_{\frac{n}{2}-1},\ldots w_{\frac{n}{2}-i+1}\}$ and $\overline{N(b_i)}=\{b_1,b_2,\ldots, b_{i-1},w_{\frac{n}{2}-i},\ldots, w_1\}$. It must be pointed out that the subgraphs induced by the sets $S_1=\{w_{\frac{n}{2}}, w_{\frac{n}{2}-1},\ldots w_{\frac{n}{2}-i+1}, b_1,b_2, \ldots, b_{i-1}\}$ and $S_2=\{b_{i+1},\ldots, b_{\frac{n}{2}}, w_{\frac{n}{2}-i},\ldots w_1\}$ are prime $P_5$-free bipartite graphs as well as $S_3=\{w_{\frac{n}{2}}, w_{\frac{n}{2}-1},\ldots w_{\frac{n}{2}-i+1}, b_{i+1},\ldots, b_{\frac{n}{2}}\}$ induces a complete bipartite graph. The result follows when considering the Seidel complementation on $b_i$ (see Figure \ref{fig:SeidelComplementationPrimeP5FreeBiparti}).
\end{prf}
\begin{cor}\label{cor:SeidelComplementationEnTempsConstant}
For a prime $P_5$-free bipartite graph (or its complement) Seidel complementation at any vertex can be performed in constant time.
\end{cor}

\ignore{\begin{cor}\label{cor:StructureComplementPrimeP5FreeBiparti}
Let $G=(V,E)$  be a prime graph of $n$ vertices.
 $\overline{G}$  is bipartite and $P_5$-free
iff the following conditions are verified:
\begin{itemize}
\item[(i)] There exists a partition of $V(G)$ into two complete sets $B$ and $W$;
\item[(ii)] The vertex-degree sequences of classes $B$ and $W$ are given by:\\
$S_B=(n/2-1,n/2,\ldots,n-2)$; $S_W=(n/2-1,n/2,\ldots,n-2)$.
\end{itemize}
\end{cor}
}
\bibliographystyle{plain}
\bibliography{P5HB}
\ignore{

}
\end{document}

%% file: P5HB-free.eepic
\setlength{\unitlength}{0.00026247in}
\begingroup\makeatletter\ifx\SetFigFont\undefined
\def\x#1#2#3#4#5#6#7\relax{\def\x{#1#2#3#4#5#6}}%
\expandafter\x\fmtname xxxxxx\relax \def\y{splain}%
\ifx\x\y   
\gdef\SetFigFont#1#2#3{%
  \ifnum #1<17\tiny\else \ifnum #1<20\small\else
  \ifnum #1<24\normalsize\else \ifnum #1<29\large\else
  \ifnum #1<34\Large\else \ifnum #1<41\LARGE\else
     \huge\fi\fi\fi\fi\fi\fi
  \csname #3\endcsname}%
\else
\gdef\SetFigFont#1#2#3{\begingroup
  \count@#1\relax \ifnum 25<\count@\count@25\fi
  \def\x{\endgroup\@setsize\SetFigFont{#2pt}}%
  \expandafter\x
    \csname \romannumeral\the\count@ pt\expandafter\endcsname
    \csname @\romannumeral\the\count@ pt\endcsname
  \csname #3\endcsname}%
\fi
\fi\endgroup
{\renewcommand{\dashlinestretch}{30}
\begin{picture}(8476,2332)(0,-10)
\put(1628,60){\makebox(0,0)[b]{\smash{{{\SetFigFont{5}{6.0}{rm}$P5$}}}}}
\put(7118,2220){\blacken\ellipse{180}{180}}
\put(7118,2220){\ellipse{180}{180}}
\put(8378,1230){\blacken\ellipse{180}{180}}
\put(8378,1230){\ellipse{180}{180}}
\put(7118,1230){\blacken\ellipse{180}{180}}
\put(7118,1230){\ellipse{180}{180}}
\put(5948,600){\blacken\ellipse{180}{180}}
\put(5948,600){\ellipse{180}{180}}
\put(7748,600){\blacken\ellipse{180}{180}}
\put(7748,600){\ellipse{180}{180}}
\put(5048,600){\blacken\ellipse{180}{180}}
\put(5048,600){\ellipse{180}{180}}
\put(5948,1500){\blacken\ellipse{180}{180}}
\put(5948,1500){\ellipse{180}{180}}
\put(5048,1500){\blacken\ellipse{180}{180}}
\put(5048,1500){\ellipse{180}{180}}
\put(5498,2220){\blacken\ellipse{180}{180}}
\put(5498,2220){\ellipse{180}{180}}
\put(3698,1050){\blacken\ellipse{180}{180}}
\put(3698,1050){\ellipse{180}{180}}
\put(2798,1050){\blacken\ellipse{180}{180}}
\put(2798,1050){\ellipse{180}{180}}
\put(1898,1050){\blacken\ellipse{180}{180}}
\put(1898,1050){\ellipse{180}{180}}
\put(998,1050){\blacken\ellipse{180}{180}}
\put(998,1050){\ellipse{180}{180}}
\put(98,1050){\blacken\ellipse{180}{180}}
\put(98,1050){\ellipse{180}{180}}
\path(7118,1230)(8378,1230)
\path(8378,2220)(8378,1230)(7748,600)
	(7118,1230)(7118,2220)
\path(5948,1500)(5048,1500)(5498,2220)
	(5948,1500)(5948,600)(5048,600)(5048,1500)
\path(3698,1050)(98,1050)
\path(3698,1050)(98,1050)
\put(7748,150){\makebox(0,0)[b]{\smash{{{\SetFigFont{5}{6.0}{rm}$Bull$}}}}}
\put(5408,150){\makebox(0,0)[b]{\smash{{{\SetFigFont{5}{6.0}{rm}$House$}}}}}
\put(8378,2220){\blacken\ellipse{180}{180}}
\put(8378,2220){\ellipse{180}{180}}
\end{picture}
}

%% file: SommetvSurH.eepic
\setlength{\unitlength}{0.00026247in}
\begingroup\makeatletter\ifx\SetFigFont\undefined
\def\x#1#2#3#4#5#6#7\relax{\def\x{#1#2#3#4#5#6}}%
\expandafter\x\fmtname xxxxxx\relax \def\y{splain}%
\ifx\x\y   
\gdef\SetFigFont#1#2#3{%
  \ifnum #1<17\tiny\else \ifnum #1<20\small\else
  \ifnum #1<24\normalsize\else \ifnum #1<29\large\else
  \ifnum #1<34\Large\else \ifnum #1<41\LARGE\else
     \huge\fi\fi\fi\fi\fi\fi
  \csname #3\endcsname}%
\else
\gdef\SetFigFont#1#2#3{\begingroup
  \count@#1\relax \ifnum 25<\count@\count@25\fi
  \def\x{\endgroup\@setsize\SetFigFont{#2pt}}%
  \expandafter\x
    \csname \romannumeral\the\count@ pt\expandafter\endcsname
    \csname @\romannumeral\the\count@ pt\endcsname
  \csname #3\endcsname}%
\fi
\fi\endgroup
{\renewcommand{\dashlinestretch}{30}
\begin{picture}(13610,9694)(0,-10)
\put(5138,7572){\makebox(0,0)[b]{\smash{{{\SetFigFont{5}{6.0}{rm}$v$}}}}}
\put(1448,8472){\blacken\ellipse{180}{180}}
\put(1448,8472){\ellipse{180}{180}}
\put(2888,8472){\blacken\ellipse{180}{180}}
\put(2888,8472){\ellipse{180}{180}}
\put(3158,7572){\blacken\ellipse{180}{180}}
\put(3158,7572){\ellipse{180}{180}}
\put(3338,8022){\blacken\ellipse{180}{180}}
\put(3338,8022){\ellipse{180}{180}}
\put(1178,7302){\arc{360}{1.5708}{3.1416}}
\put(1178,8742){\arc{360}{3.1416}{4.7124}}
\put(1808,8742){\arc{360}{4.7124}{6.2832}}
\put(1808,7302){\arc{360}{0}{1.5708}}
\path(998,7302)(998,8742)
\path(1178,8922)(1808,8922)
\path(1988,8742)(1988,7302)
\path(1808,7122)(1178,7122)
\put(2798,7302){\arc{360}{1.5708}{3.1416}}
\put(2798,8742){\arc{360}{3.1416}{4.7124}}
\put(3428,8742){\arc{360}{4.7124}{6.2832}}
\put(3428,7302){\arc{360}{0}{1.5708}}
\path(2618,7302)(2618,8742)
\path(2798,8922)(3428,8922)
\path(3608,8742)(3608,7302)
\path(3428,7122)(2798,7122)
\path(98,8022)(1448,8472)(2888,8472)
	(3338,8022)(3158,7572)
\put(1538,9372){\makebox(0,0)[b]{\smash{{{\SetFigFont{5}{6.0}{rm}$N(v)$}}}}}
\put(3068,9372){\makebox(0,0)[b]{\smash{{{\SetFigFont{5}{6.0}{rm}$\overline{N(v)}$}}}}}
\put(98,7572){\makebox(0,0)[b]{\smash{{{\SetFigFont{5}{6.0}{rm}$v$}}}}}
\put(10088,1002){\blacken\ellipse{180}{180}}
\put(10088,1002){\ellipse{180}{180}}
\put(11438,1452){\blacken\ellipse{180}{180}}
\put(11438,1452){\ellipse{180}{180}}
\put(12878,1452){\blacken\ellipse{180}{180}}
\put(12878,1452){\ellipse{180}{180}}
\put(11438,642){\blacken\ellipse{180}{180}}
\put(11438,642){\ellipse{180}{180}}
\put(13328,1002){\blacken\ellipse{180}{180}}
\put(13328,1002){\ellipse{180}{180}}
\put(6488,732){\blacken\ellipse{180}{180}}
\put(6488,732){\ellipse{180}{180}}
\put(6848,372){\blacken\ellipse{180}{180}}
\put(6848,372){\ellipse{180}{180}}
\put(7928,1362){\blacken\ellipse{180}{180}}
\put(7928,1362){\ellipse{180}{180}}
\put(6488,1362){\blacken\ellipse{180}{180}}
\put(6488,1362){\ellipse{180}{180}}
\put(5138,912){\blacken\ellipse{180}{180}}
\put(5138,912){\ellipse{180}{180}}
\put(188,1002){\blacken\ellipse{180}{180}}
\put(188,1002){\ellipse{180}{180}}
\put(1538,1452){\blacken\ellipse{180}{180}}
\put(1538,1452){\ellipse{180}{180}}
\put(3068,1452){\blacken\ellipse{180}{180}}
\put(3068,1452){\ellipse{180}{180}}
\put(1538,552){\blacken\ellipse{180}{180}}
\put(1538,552){\ellipse{180}{180}}
\put(3068,552){\blacken\ellipse{180}{180}}
\put(3068,552){\ellipse{180}{180}}
\put(10088,4602){\blacken\ellipse{180}{180}}
\put(10088,4602){\ellipse{180}{180}}
\put(11438,5052){\blacken\ellipse{180}{180}}
\put(11438,5052){\ellipse{180}{180}}
\put(12878,5052){\blacken\ellipse{180}{180}}
\put(12878,5052){\ellipse{180}{180}}
\put(12968,4152){\blacken\ellipse{180}{180}}
\put(12968,4152){\ellipse{180}{180}}
\put(11438,4152){\blacken\ellipse{180}{180}}
\put(11438,4152){\ellipse{180}{180}}
\put(5138,4602){\blacken\ellipse{180}{180}}
\put(5138,4602){\ellipse{180}{180}}
\put(6488,5052){\blacken\ellipse{180}{180}}
\put(6488,5052){\ellipse{180}{180}}
\put(7928,5052){\blacken\ellipse{180}{180}}
\put(7928,5052){\ellipse{180}{180}}
\put(6848,4422){\blacken\ellipse{180}{180}}
\put(6848,4422){\ellipse{180}{180}}
\put(6398,3882){\blacken\ellipse{180}{180}}
\put(6398,3882){\ellipse{180}{180}}
\put(3518,4512){\blacken\ellipse{180}{180}}
\put(3518,4512){\ellipse{180}{180}}
\put(188,4602){\blacken\ellipse{180}{180}}
\put(188,4602){\ellipse{180}{180}}
\put(1538,5052){\blacken\ellipse{180}{180}}
\put(1538,5052){\ellipse{180}{180}}
\put(2978,5052){\blacken\ellipse{180}{180}}
\put(2978,5052){\ellipse{180}{180}}
\put(2978,4062){\blacken\ellipse{180}{180}}
\put(2978,4062){\ellipse{180}{180}}
\put(10088,8022){\blacken\ellipse{180}{180}}
\put(10088,8022){\ellipse{180}{180}}
\put(11438,8472){\blacken\ellipse{180}{180}}
\put(11438,8472){\ellipse{180}{180}}
\put(12878,8472){\blacken\ellipse{180}{180}}
\put(12878,8472){\ellipse{180}{180}}
\put(11528,7572){\blacken\ellipse{180}{180}}
\put(11528,7572){\ellipse{180}{180}}
\put(12878,7572){\blacken\ellipse{180}{180}}
\put(12878,7572){\ellipse{180}{180}}
\put(5138,8022){\blacken\ellipse{180}{180}}
\put(5138,8022){\ellipse{180}{180}}
\put(6488,8472){\blacken\ellipse{180}{180}}
\put(6488,8472){\ellipse{180}{180}}
\put(7928,8472){\blacken\ellipse{180}{180}}
\put(7928,8472){\ellipse{180}{180}}
\put(6578,7572){\blacken\ellipse{180}{180}}
\put(6578,7572){\ellipse{180}{180}}
\put(8378,8022){\blacken\ellipse{180}{180}}
\put(8378,8022){\ellipse{180}{180}}
\path(11438,552)(13328,1002)(12878,1452)
	(11438,642)(10088,1002)(11438,1452)(12878,1452)
\put(11168,282){\arc{360}{1.5708}{3.1416}}
\put(11168,1722){\arc{360}{3.1416}{4.7124}}
\put(11798,1722){\arc{360}{4.7124}{6.2832}}
\put(11798,282){\arc{360}{0}{1.5708}}
\path(10988,282)(10988,1722)
\path(11168,1902)(11798,1902)
\path(11978,1722)(11978,282)
\path(11798,102)(11168,102)
\put(12788,282){\arc{360}{1.5708}{3.1416}}
\put(12788,1722){\arc{360}{3.1416}{4.7124}}
\put(13418,1722){\arc{360}{4.7124}{6.2832}}
\put(13418,282){\arc{360}{0}{1.5708}}
\path(12608,282)(12608,1722)
\path(12788,1902)(13418,1902)
\path(13598,1722)(13598,282)
\path(13418,102)(12788,102)
\path(5138,912)(6848,372)(7928,1362)
\path(7928,1362)(6488,1362)(5138,912)
	(6488,732)(6488,1272)
\put(7838,192){\arc{360}{1.5708}{3.1416}}
\put(7838,1632){\arc{360}{3.1416}{4.7124}}
\put(8468,1632){\arc{360}{4.7124}{6.2832}}
\put(8468,192){\arc{360}{0}{1.5708}}
\path(7658,192)(7658,1632)
\path(7838,1812)(8468,1812)
\path(8648,1632)(8648,192)
\path(8468,12)(7838,12)
\put(6218,192){\arc{360}{1.5708}{3.1416}}
\put(6218,1632){\arc{360}{3.1416}{4.7124}}
\put(6848,1632){\arc{360}{4.7124}{6.2832}}
\put(6848,192){\arc{360}{0}{1.5708}}
\path(6038,192)(6038,1632)
\path(6218,1812)(6848,1812)
\path(7028,1632)(7028,192)
\path(6848,12)(6218,12)
\path(1538,1362)(1538,552)(3068,552)
	(3068,1452)(1538,1452)(188,1002)(1538,552)
\put(1268,282){\arc{360}{1.5708}{3.1416}}
\put(1268,1722){\arc{360}{3.1416}{4.7124}}
\put(1898,1722){\arc{360}{4.7124}{6.2832}}
\put(1898,282){\arc{360}{0}{1.5708}}
\path(1088,282)(1088,1722)
\path(1268,1902)(1898,1902)
\path(2078,1722)(2078,282)
\path(1898,102)(1268,102)
\put(2888,282){\arc{360}{1.5708}{3.1416}}
\put(2888,1722){\arc{360}{3.1416}{4.7124}}
\put(3518,1722){\arc{360}{4.7124}{6.2832}}
\put(3518,282){\arc{360}{0}{1.5708}}
\path(2708,282)(2708,1722)
\path(2888,1902)(3518,1902)
\path(3698,1722)(3698,282)
\path(3518,102)(2888,102)
\path(11438,4152)(10088,4602)
\path(12878,5052)(11528,5052)
\path(10088,4602)(11438,5052)(11438,4152)(12968,4152)
\put(11168,3882){\arc{360}{1.5708}{3.1416}}
\put(11168,5322){\arc{360}{3.1416}{4.7124}}
\put(11798,5322){\arc{360}{4.7124}{6.2832}}
\put(11798,3882){\arc{360}{0}{1.5708}}
\path(10988,3882)(10988,5322)
\path(11168,5502)(11798,5502)
\path(11978,5322)(11978,3882)
\path(11798,3702)(11168,3702)
\put(12788,3882){\arc{360}{1.5708}{3.1416}}
\put(12788,5322){\arc{360}{3.1416}{4.7124}}
\put(13418,5322){\arc{360}{4.7124}{6.2832}}
\put(13418,3882){\arc{360}{0}{1.5708}}
\path(12608,3882)(12608,5322)
\path(12788,5502)(13418,5502)
\path(13598,5322)(13598,3882)
\path(13418,3702)(12788,3702)
\path(6488,5052)(6848,4422)(5138,4602)
\path(7928,5052)(6488,5052)(5228,4602)(6398,3882)
\put(6218,3882){\arc{360}{1.5708}{3.1416}}
\put(6218,5322){\arc{360}{3.1416}{4.7124}}
\put(6848,5322){\arc{360}{4.7124}{6.2832}}
\put(6848,3882){\arc{360}{0}{1.5708}}
\path(6038,3882)(6038,5322)
\path(6218,5502)(6848,5502)
\path(7028,5322)(7028,3882)
\path(6848,3702)(6218,3702)
\put(7838,3882){\arc{360}{1.5708}{3.1416}}
\put(7838,5322){\arc{360}{3.1416}{4.7124}}
\put(8468,5322){\arc{360}{4.7124}{6.2832}}
\put(8468,3882){\arc{360}{0}{1.5708}}
\path(7658,3882)(7658,5322)
\path(7838,5502)(8468,5502)
\path(8648,5322)(8648,3882)
\path(8468,3702)(7838,3702)
\path(1538,5052)(2978,4062)
\path(188,4602)(1538,5052)(2978,5052)
	(2978,4062)(3518,4512)
\put(1268,3882){\arc{360}{1.5708}{3.1416}}
\put(1268,5322){\arc{360}{3.1416}{4.7124}}
\put(1898,5322){\arc{360}{4.7124}{6.2832}}
\put(1898,3882){\arc{360}{0}{1.5708}}
\path(1088,3882)(1088,5322)
\path(1268,5502)(1898,5502)
\path(2078,5322)(2078,3882)
\path(1898,3702)(1268,3702)
\put(2888,3882){\arc{360}{1.5708}{3.1416}}
\put(2888,5322){\arc{360}{3.1416}{4.7124}}
\put(3518,5322){\arc{360}{4.7124}{6.2832}}
\put(3518,3882){\arc{360}{0}{1.5708}}
\path(2708,3882)(2708,5322)
\path(2888,5502)(3518,5502)
\path(3698,5322)(3698,3882)
\path(3518,3702)(2888,3702)
\path(12878,8472)(11438,8472)(10088,8022)
	(11528,7572)(12878,7572)
\put(11168,7302){\arc{360}{1.5708}{3.1416}}
\put(11168,8742){\arc{360}{3.1416}{4.7124}}
\put(11798,8742){\arc{360}{4.7124}{6.2832}}
\put(11798,7302){\arc{360}{0}{1.5708}}
\path(10988,7302)(10988,8742)
\path(11168,8922)(11798,8922)
\path(11978,8742)(11978,7302)
\path(11798,7122)(11168,7122)
\put(12788,7302){\arc{360}{1.5708}{3.1416}}
\put(12788,8742){\arc{360}{3.1416}{4.7124}}
\put(13418,8742){\arc{360}{4.7124}{6.2832}}
\put(13418,7302){\arc{360}{0}{1.5708}}
\path(12608,7302)(12608,8742)
\path(12788,8922)(13418,8922)
\path(13598,8742)(13598,7302)
\path(13418,7122)(12788,7122)
\path(6578,7572)(5138,8022)(6488,8472)
	(7928,8472)(8378,8022)
\put(6218,7302){\arc{360}{1.5708}{3.1416}}
\put(6218,8742){\arc{360}{3.1416}{4.7124}}
\put(6848,8742){\arc{360}{4.7124}{6.2832}}
\put(6848,7302){\arc{360}{0}{1.5708}}
\path(6038,7302)(6038,8742)
\path(6218,8922)(6848,8922)
\path(7028,8742)(7028,7302)
\path(6848,7122)(6218,7122)
\put(7838,7302){\arc{360}{1.5708}{3.1416}}
\put(7838,8742){\arc{360}{3.1416}{4.7124}}
\put(8468,8742){\arc{360}{4.7124}{6.2832}}
\put(8468,7302){\arc{360}{0}{1.5708}}
\path(7658,7302)(7658,8742)
\path(7838,8922)(8468,8922)
\path(8648,8742)(8648,7302)
\path(8468,7122)(7838,7122)
\put(11528,2352){\makebox(0,0)[b]{\smash{{{\SetFigFont{5}{6.0}{rm}$N(v)$}}}}}
\put(13058,2352){\makebox(0,0)[b]{\smash{{{\SetFigFont{5}{6.0}{rm}$\overline{N(v)}$}}}}}
\put(10088,552){\makebox(0,0)[b]{\smash{{{\SetFigFont{5}{6.0}{rm}$v$}}}}}
\put(5138,462){\makebox(0,0)[b]{\smash{{{\SetFigFont{5}{6.0}{rm}$v$}}}}}
\put(8108,2262){\makebox(0,0)[b]{\smash{{{\SetFigFont{5}{6.0}{rm}$\overline{N(v)}$}}}}}
\put(6578,2262){\makebox(0,0)[b]{\smash{{{\SetFigFont{5}{6.0}{rm}$N(v)$}}}}}
\put(1628,2352){\makebox(0,0)[b]{\smash{{{\SetFigFont{5}{6.0}{rm}$N(v)$}}}}}
\put(3158,2352){\makebox(0,0)[b]{\smash{{{\SetFigFont{5}{6.0}{rm}$\overline{N(v)}$}}}}}
\put(188,552){\makebox(0,0)[b]{\smash{{{\SetFigFont{5}{6.0}{rm}$v$}}}}}
\put(11528,5952){\makebox(0,0)[b]{\smash{{{\SetFigFont{5}{6.0}{rm}$N(v)$}}}}}
\put(13058,5952){\makebox(0,0)[b]{\smash{{{\SetFigFont{5}{6.0}{rm}$\overline{N(v)}$}}}}}
\put(10088,4152){\makebox(0,0)[b]{\smash{{{\SetFigFont{5}{6.0}{rm}$v$}}}}}
\put(6578,5952){\makebox(0,0)[b]{\smash{{{\SetFigFont{5}{6.0}{rm}$N(v)$}}}}}
\put(8108,5952){\makebox(0,0)[b]{\smash{{{\SetFigFont{5}{6.0}{rm}$\overline{N(v)}$}}}}}
\put(5138,4152){\makebox(0,0)[b]{\smash{{{\SetFigFont{5}{6.0}{rm}$v$}}}}}
\put(1628,5952){\makebox(0,0)[b]{\smash{{{\SetFigFont{5}{6.0}{rm}$N(v)$}}}}}
\put(3158,5952){\makebox(0,0)[b]{\smash{{{\SetFigFont{5}{6.0}{rm}$\overline{N(v)}$}}}}}
\put(188,4152){\makebox(0,0)[b]{\smash{{{\SetFigFont{5}{6.0}{rm}$v$}}}}}
\put(11528,9372){\makebox(0,0)[b]{\smash{{{\SetFigFont{5}{6.0}{rm}$N(v)$}}}}}
\put(13058,9372){\makebox(0,0)[b]{\smash{{{\SetFigFont{5}{6.0}{rm}$\overline{N(v)}$}}}}}
\put(10088,7572){\makebox(0,0)[b]{\smash{{{\SetFigFont{5}{6.0}{rm}$v$}}}}}
\put(6578,9372){\makebox(0,0)[b]{\smash{{{\SetFigFont{5}{6.0}{rm}$N(v)$}}}}}
\put(8108,9372){\makebox(0,0)[b]{\smash{{{\SetFigFont{5}{6.0}{rm}$\overline{N(v)}$}}}}}
\put(98,8022){\blacken\ellipse{180}{180}}
\put(98,8022){\ellipse{180}{180}}
\end{picture}
}

%% file: SommetvPasSurP5.eepic
\setlength{\unitlength}{0.00026247in}
\begingroup\makeatletter\ifx\SetFigFont\undefined
\def\x#1#2#3#4#5#6#7\relax{\def\x{#1#2#3#4#5#6}}%
\expandafter\x\fmtname xxxxxx\relax \def\y{splain}%
\ifx\x\y   
\gdef\SetFigFont#1#2#3{%
  \ifnum #1<17\tiny\else \ifnum #1<20\small\else
  \ifnum #1<24\normalsize\else \ifnum #1<29\large\else
  \ifnum #1<34\Large\else \ifnum #1<41\LARGE\else
     \huge\fi\fi\fi\fi\fi\fi
  \csname #3\endcsname}%
\else
\gdef\SetFigFont#1#2#3{\begingroup
  \count@#1\relax \ifnum 25<\count@\count@25\fi
  \def\x{\endgroup\@setsize\SetFigFont{#2pt}}%
  \expandafter\x
    \csname \romannumeral\the\count@ pt\expandafter\endcsname
    \csname @\romannumeral\the\count@ pt\endcsname
  \csname #3\endcsname}%
\fi
\fi\endgroup
{\renewcommand{\dashlinestretch}{30}
\begin{picture}(13610,9694)(0,-10)
\put(3068,9372){\makebox(0,0)[b]{\smash{{{\SetFigFont{5}{6.0}{rm}$\overline{N(v)}$}}}}}
\put(188,912){\blacken\ellipse{180}{180}}
\put(188,912){\ellipse{180}{180}}
\put(1538,642){\blacken\ellipse{180}{180}}
\put(1538,642){\ellipse{180}{180}}
\put(2978,1362){\blacken\ellipse{180}{180}}
\put(2978,1362){\ellipse{180}{180}}
\put(2978,372){\blacken\ellipse{180}{180}}
\put(2978,372){\ellipse{180}{180}}
\put(1538,1272){\blacken\ellipse{180}{180}}
\put(1538,1272){\ellipse{180}{180}}
\put(1268,192){\arc{360}{1.5708}{3.1416}}
\put(1268,1632){\arc{360}{3.1416}{4.7124}}
\put(1898,1632){\arc{360}{4.7124}{6.2832}}
\put(1898,192){\arc{360}{0}{1.5708}}
\path(1088,192)(1088,1632)
\path(1268,1812)(1898,1812)
\path(2078,1632)(2078,192)
\path(1898,12)(1268,12)
\put(2888,192){\arc{360}{1.5708}{3.1416}}
\put(2888,1632){\arc{360}{3.1416}{4.7124}}
\put(3518,1632){\arc{360}{4.7124}{6.2832}}
\put(3518,192){\arc{360}{0}{1.5708}}
\path(2708,192)(2708,1632)
\path(2888,1812)(3518,1812)
\path(3698,1632)(3698,192)
\path(3518,12)(2888,12)
\put(1628,2262){\makebox(0,0)[b]{\smash{{{\SetFigFont{5}{6.0}{rm}$N(v)$}}}}}
\put(3158,2262){\makebox(0,0)[b]{\smash{{{\SetFigFont{5}{6.0}{rm}$\overline{N(v)}$}}}}}
\put(188,462){\makebox(0,0)[b]{\smash{{{\SetFigFont{5}{6.0}{rm}$v$}}}}}
\put(11348,1272){\blacken\ellipse{180}{180}}
\put(11348,1272){\ellipse{180}{180}}
\put(12788,372){\blacken\ellipse{180}{180}}
\put(12788,372){\ellipse{180}{180}}
\put(12788,1362){\blacken\ellipse{180}{180}}
\put(12788,1362){\ellipse{180}{180}}
\put(11348,642){\blacken\ellipse{180}{180}}
\put(11348,642){\ellipse{180}{180}}
\put(9998,912){\blacken\ellipse{180}{180}}
\put(9998,912){\ellipse{180}{180}}
\put(13328,822){\blacken\ellipse{180}{180}}
\put(13328,822){\ellipse{180}{180}}
\put(12698,192){\arc{360}{1.5708}{3.1416}}
\put(12698,1632){\arc{360}{3.1416}{4.7124}}
\put(13328,1632){\arc{360}{4.7124}{6.2832}}
\put(13328,192){\arc{360}{0}{1.5708}}
\path(12518,192)(12518,1632)
\path(12698,1812)(13328,1812)
\path(13508,1632)(13508,192)
\path(13328,12)(12698,12)
\put(11078,192){\arc{360}{1.5708}{3.1416}}
\put(11078,1632){\arc{360}{3.1416}{4.7124}}
\put(11708,1632){\arc{360}{4.7124}{6.2832}}
\put(11708,192){\arc{360}{0}{1.5708}}
\path(10898,192)(10898,1632)
\path(11078,1812)(11708,1812)
\path(11888,1632)(11888,192)
\path(11708,12)(11078,12)
\put(9998,462){\makebox(0,0)[b]{\smash{{{\SetFigFont{5}{6.0}{rm}$v$}}}}}
\put(12968,2262){\makebox(0,0)[b]{\smash{{{\SetFigFont{5}{6.0}{rm}$\overline{N(v)}$}}}}}
\put(11438,2262){\makebox(0,0)[b]{\smash{{{\SetFigFont{5}{6.0}{rm}$N(v)$}}}}}
\put(6488,1272){\blacken\ellipse{180}{180}}
\put(6488,1272){\ellipse{180}{180}}
\put(7928,372){\blacken\ellipse{180}{180}}
\put(7928,372){\ellipse{180}{180}}
\put(7928,1362){\blacken\ellipse{180}{180}}
\put(7928,1362){\ellipse{180}{180}}
\put(6488,642){\blacken\ellipse{180}{180}}
\put(6488,642){\ellipse{180}{180}}
\put(5138,912){\blacken\ellipse{180}{180}}
\put(5138,912){\ellipse{180}{180}}
\put(8468,822){\blacken\ellipse{180}{180}}
\put(8468,822){\ellipse{180}{180}}
\put(7838,192){\arc{360}{1.5708}{3.1416}}
\put(7838,1632){\arc{360}{3.1416}{4.7124}}
\put(8468,1632){\arc{360}{4.7124}{6.2832}}
\put(8468,192){\arc{360}{0}{1.5708}}
\path(7658,192)(7658,1632)
\path(7838,1812)(8468,1812)
\path(8648,1632)(8648,192)
\path(8468,12)(7838,12)
\put(6218,192){\arc{360}{1.5708}{3.1416}}
\put(6218,1632){\arc{360}{3.1416}{4.7124}}
\put(6848,1632){\arc{360}{4.7124}{6.2832}}
\put(6848,192){\arc{360}{0}{1.5708}}
\path(6038,192)(6038,1632)
\path(6218,1812)(6848,1812)
\path(7028,1632)(7028,192)
\path(6848,12)(6218,12)
\put(5138,462){\makebox(0,0)[b]{\smash{{{\SetFigFont{5}{6.0}{rm}$v$}}}}}
\put(8108,2262){\makebox(0,0)[b]{\smash{{{\SetFigFont{5}{6.0}{rm}$\overline{N(v)}$}}}}}
\put(6578,2262){\makebox(0,0)[b]{\smash{{{\SetFigFont{5}{6.0}{rm}$N(v)$}}}}}
\put(2888,8202){\blacken\ellipse{180}{180}}
\put(2888,8202){\ellipse{180}{180}}
\put(2888,7752){\blacken\ellipse{180}{180}}
\put(2888,7752){\ellipse{180}{180}}
\put(2888,8652){\blacken\ellipse{180}{180}}
\put(2888,8652){\ellipse{180}{180}}
\put(1448,8472){\blacken\ellipse{180}{180}}
\put(1448,8472){\ellipse{180}{180}}
\put(98,8022){\blacken\ellipse{180}{180}}
\put(98,8022){\ellipse{180}{180}}
\put(2888,7302){\blacken\ellipse{180}{180}}
\put(2888,7302){\ellipse{180}{180}}
\put(2798,7302){\arc{360}{1.5708}{3.1416}}
\put(2798,8742){\arc{360}{3.1416}{4.7124}}
\put(3428,8742){\arc{360}{4.7124}{6.2832}}
\put(3428,7302){\arc{360}{0}{1.5708}}
\path(2618,7302)(2618,8742)
\path(2798,8922)(3428,8922)
\path(3608,8742)(3608,7302)
\path(3428,7122)(2798,7122)
\put(1178,7302){\arc{360}{1.5708}{3.1416}}
\put(1178,8742){\arc{360}{3.1416}{4.7124}}
\put(1808,8742){\arc{360}{4.7124}{6.2832}}
\put(1808,7302){\arc{360}{0}{1.5708}}
\path(998,7302)(998,8742)
\path(1178,8922)(1808,8922)
\path(1988,8742)(1988,7302)
\path(1808,7122)(1178,7122)
\path(2888,7302)(2888,8562)(1448,8472)(188,8022)
\put(98,7572){\makebox(0,0)[b]{\smash{{{\SetFigFont{5}{6.0}{rm}$v$}}}}}
\put(1538,9372){\makebox(0,0)[b]{\smash{{{\SetFigFont{5}{6.0}{rm}$N(v)$}}}}}
\put(11438,4872){\blacken\ellipse{180}{180}}
\put(11438,4872){\ellipse{180}{180}}
\put(12878,3972){\blacken\ellipse{180}{180}}
\put(12878,3972){\ellipse{180}{180}}
\put(12878,4962){\blacken\ellipse{180}{180}}
\put(12878,4962){\ellipse{180}{180}}
\put(11438,4242){\blacken\ellipse{180}{180}}
\put(11438,4242){\ellipse{180}{180}}
\put(10088,4512){\blacken\ellipse{180}{180}}
\put(10088,4512){\ellipse{180}{180}}
\put(13418,4422){\blacken\ellipse{180}{180}}
\put(13418,4422){\ellipse{180}{180}}
\put(6398,4872){\blacken\ellipse{180}{180}}
\put(6398,4872){\ellipse{180}{180}}
\put(7838,3972){\blacken\ellipse{180}{180}}
\put(7838,3972){\ellipse{180}{180}}
\put(7838,4962){\blacken\ellipse{180}{180}}
\put(7838,4962){\ellipse{180}{180}}
\put(6398,4242){\blacken\ellipse{180}{180}}
\put(6398,4242){\ellipse{180}{180}}
\put(5048,4512){\blacken\ellipse{180}{180}}
\put(5048,4512){\ellipse{180}{180}}
\put(8378,4422){\blacken\ellipse{180}{180}}
\put(8378,4422){\ellipse{180}{180}}
\put(1538,4962){\blacken\ellipse{180}{180}}
\put(1538,4962){\ellipse{180}{180}}
\put(2978,4062){\blacken\ellipse{180}{180}}
\put(2978,4062){\ellipse{180}{180}}
\put(2978,5052){\blacken\ellipse{180}{180}}
\put(2978,5052){\ellipse{180}{180}}
\put(1538,4332){\blacken\ellipse{180}{180}}
\put(1538,4332){\ellipse{180}{180}}
\put(188,4602){\blacken\ellipse{180}{180}}
\put(188,4602){\ellipse{180}{180}}
\put(3518,4512){\blacken\ellipse{180}{180}}
\put(3518,4512){\ellipse{180}{180}}
\put(13418,8472){\blacken\ellipse{180}{180}}
\put(13418,8472){\ellipse{180}{180}}
\put(12878,7842){\blacken\ellipse{180}{180}}
\put(12878,7842){\ellipse{180}{180}}
\put(12878,8472){\blacken\ellipse{180}{180}}
\put(12878,8472){\ellipse{180}{180}}
\put(11528,8112){\blacken\ellipse{180}{180}}
\put(11528,8112){\ellipse{180}{180}}
\put(10088,8112){\blacken\ellipse{180}{180}}
\put(10088,8112){\ellipse{180}{180}}
\put(13418,7842){\blacken\ellipse{180}{180}}
\put(13418,7842){\ellipse{180}{180}}
\put(7838,8292){\blacken\ellipse{180}{180}}
\put(7838,8292){\ellipse{180}{180}}
\put(7838,7842){\blacken\ellipse{180}{180}}
\put(7838,7842){\ellipse{180}{180}}
\put(7838,8742){\blacken\ellipse{180}{180}}
\put(7838,8742){\ellipse{180}{180}}
\put(6398,8112){\blacken\ellipse{180}{180}}
\put(6398,8112){\ellipse{180}{180}}
\put(5048,8112){\blacken\ellipse{180}{180}}
\put(5048,8112){\ellipse{180}{180}}
\put(7838,7392){\blacken\ellipse{180}{180}}
\put(7838,7392){\ellipse{180}{180}}
\path(9998,912)(11348,1272)(12788,1362)
	(13328,822)(12788,372)(11348,642)(9998,912)
\path(7928,372)(6488,642)(5138,912)
	(6398,1272)(7928,1362)(8468,822)(6488,642)
\path(2978,372)(3518,822)(1538,642)
	(3068,1362)(1538,1272)(188,912)(1538,642)
\path(11438,4872)(10088,4512)(11438,4242)
\path(12878,4962)(11438,4872)(13418,4422)
	(11438,4242)(12878,3972)
\path(6398,4872)(5048,4512)(6398,4242)
\path(7838,4962)(6398,4872)(6398,4242)
	(8378,4422)(7838,3972)
\path(1538,4962)(188,4602)(1538,4332)
\path(1538,4332)(1538,4962)(2978,5052)
	(3518,4512)(2978,4062)
\put(12788,3792){\arc{360}{1.5708}{3.1416}}
\put(12788,5232){\arc{360}{3.1416}{4.7124}}
\put(13418,5232){\arc{360}{4.7124}{6.2832}}
\put(13418,3792){\arc{360}{0}{1.5708}}
\path(12608,3792)(12608,5232)
\path(12788,5412)(13418,5412)
\path(13598,5232)(13598,3792)
\path(13418,3612)(12788,3612)
\put(11168,3792){\arc{360}{1.5708}{3.1416}}
\put(11168,5232){\arc{360}{3.1416}{4.7124}}
\put(11798,5232){\arc{360}{4.7124}{6.2832}}
\put(11798,3792){\arc{360}{0}{1.5708}}
\path(10988,3792)(10988,5232)
\path(11168,5412)(11798,5412)
\path(11978,5232)(11978,3792)
\path(11798,3612)(11168,3612)
\put(7748,3792){\arc{360}{1.5708}{3.1416}}
\put(7748,5232){\arc{360}{3.1416}{4.7124}}
\put(8378,5232){\arc{360}{4.7124}{6.2832}}
\put(8378,3792){\arc{360}{0}{1.5708}}
\path(7568,3792)(7568,5232)
\path(7748,5412)(8378,5412)
\path(8558,5232)(8558,3792)
\path(8378,3612)(7748,3612)
\put(6128,3792){\arc{360}{1.5708}{3.1416}}
\put(6128,5232){\arc{360}{3.1416}{4.7124}}
\put(6758,5232){\arc{360}{4.7124}{6.2832}}
\put(6758,3792){\arc{360}{0}{1.5708}}
\path(5948,3792)(5948,5232)
\path(6128,5412)(6758,5412)
\path(6938,5232)(6938,3792)
\path(6758,3612)(6128,3612)
\put(2888,3882){\arc{360}{1.5708}{3.1416}}
\put(2888,5322){\arc{360}{3.1416}{4.7124}}
\put(3518,5322){\arc{360}{4.7124}{6.2832}}
\put(3518,3882){\arc{360}{0}{1.5708}}
\path(2708,3882)(2708,5322)
\path(2888,5502)(3518,5502)
\path(3698,5322)(3698,3882)
\path(3518,3702)(2888,3702)
\put(1268,3882){\arc{360}{1.5708}{3.1416}}
\put(1268,5322){\arc{360}{3.1416}{4.7124}}
\put(1898,5322){\arc{360}{4.7124}{6.2832}}
\put(1898,3882){\arc{360}{0}{1.5708}}
\path(1088,3882)(1088,5322)
\path(1268,5502)(1898,5502)
\path(2078,5322)(2078,3882)
\path(1898,3702)(1268,3702)
\path(11528,8112)(10088,8112)
\path(13418,8472)(12878,8472)(11528,8112)
	(12878,7842)(13418,7842)
\put(12788,7392){\arc{360}{1.5708}{3.1416}}
\put(12788,8832){\arc{360}{3.1416}{4.7124}}
\put(13418,8832){\arc{360}{4.7124}{6.2832}}
\put(13418,7392){\arc{360}{0}{1.5708}}
\path(12608,7392)(12608,8832)
\path(12788,9012)(13418,9012)
\path(13598,8832)(13598,7392)
\path(13418,7212)(12788,7212)
\put(11168,7392){\arc{360}{1.5708}{3.1416}}
\put(11168,8832){\arc{360}{3.1416}{4.7124}}
\put(11798,8832){\arc{360}{4.7124}{6.2832}}
\put(11798,7392){\arc{360}{0}{1.5708}}
\path(10988,7392)(10988,8832)
\path(11168,9012)(11798,9012)
\path(11978,8832)(11978,7392)
\path(11798,7212)(11168,7212)
\path(5048,8112)(6398,8112)
\path(7838,8742)(6398,8112)(7838,8292)(7838,7392)
\put(7748,7392){\arc{360}{1.5708}{3.1416}}
\put(7748,8832){\arc{360}{3.1416}{4.7124}}
\put(8378,8832){\arc{360}{4.7124}{6.2832}}
\put(8378,7392){\arc{360}{0}{1.5708}}
\path(7568,7392)(7568,8832)
\path(7748,9012)(8378,9012)
\path(8558,8832)(8558,7392)
\path(8378,7212)(7748,7212)
\put(6128,7392){\arc{360}{1.5708}{3.1416}}
\put(6128,8832){\arc{360}{3.1416}{4.7124}}
\put(6758,8832){\arc{360}{4.7124}{6.2832}}
\put(6758,7392){\arc{360}{0}{1.5708}}
\path(5948,7392)(5948,8832)
\path(6128,9012)(6758,9012)
\path(6938,8832)(6938,7392)
\path(6758,7212)(6128,7212)
\put(10088,4062){\makebox(0,0)[b]{\smash{{{\SetFigFont{5}{6.0}{rm}$v$}}}}}
\put(13058,5862){\makebox(0,0)[b]{\smash{{{\SetFigFont{5}{6.0}{rm}$\overline{N(v)}$}}}}}
\put(11528,5862){\makebox(0,0)[b]{\smash{{{\SetFigFont{5}{6.0}{rm}$N(v)$}}}}}
\put(5048,4062){\makebox(0,0)[b]{\smash{{{\SetFigFont{5}{6.0}{rm}$v$}}}}}
\put(8018,5862){\makebox(0,0)[b]{\smash{{{\SetFigFont{5}{6.0}{rm}$\overline{N(v)}$}}}}}
\put(6488,5862){\makebox(0,0)[b]{\smash{{{\SetFigFont{5}{6.0}{rm}$N(v)$}}}}}
\put(188,4152){\makebox(0,0)[b]{\smash{{{\SetFigFont{5}{6.0}{rm}$v$}}}}}
\put(3158,5952){\makebox(0,0)[b]{\smash{{{\SetFigFont{5}{6.0}{rm}$\overline{N(v)}$}}}}}
\put(1628,5952){\makebox(0,0)[b]{\smash{{{\SetFigFont{5}{6.0}{rm}$N(v)$}}}}}
\put(13058,9372){\makebox(0,0)[b]{\smash{{{\SetFigFont{5}{6.0}{rm}$\overline{N(v)}$}}}}}
\put(8018,9372){\makebox(0,0)[b]{\smash{{{\SetFigFont{5}{6.0}{rm}$\overline{N(v)}$}}}}}
\put(10088,7662){\makebox(0,0)[b]{\smash{{{\SetFigFont{5}{6.0}{rm}$v$}}}}}
\put(11528,9462){\makebox(0,0)[b]{\smash{{{\SetFigFont{5}{6.0}{rm}$N(v)$}}}}}
\put(5048,7662){\makebox(0,0)[b]{\smash{{{\SetFigFont{5}{6.0}{rm}$v$}}}}}
\put(6488,9462){\makebox(0,0)[b]{\smash{{{\SetFigFont{5}{6.0}{rm}$N(v)$}}}}}
\put(3518,822){\blacken\ellipse{180}{180}}
\put(3518,822){\ellipse{180}{180}}
\end{picture}
}

%% file: SommetvPasSurBull.eepic
\setlength{\unitlength}{0.00026247in}
\begingroup\makeatletter\ifx\SetFigFont\undefined
\def\x#1#2#3#4#5#6#7\relax{\def\x{#1#2#3#4#5#6}}%
\expandafter\x\fmtname xxxxxx\relax \def\y{splain}%
\ifx\x\y   
\gdef\SetFigFont#1#2#3{%
  \ifnum #1<17\tiny\else \ifnum #1<20\small\else
  \ifnum #1<24\normalsize\else \ifnum #1<29\large\else
  \ifnum #1<34\Large\else \ifnum #1<41\LARGE\else
     \huge\fi\fi\fi\fi\fi\fi
  \csname #3\endcsname}%
\else
\gdef\SetFigFont#1#2#3{\begingroup
  \count@#1\relax \ifnum 25<\count@\count@25\fi
  \def\x{\endgroup\@setsize\SetFigFont{#2pt}}%
  \expandafter\x
    \csname \romannumeral\the\count@ pt\expandafter\endcsname
    \csname @\romannumeral\the\count@ pt\endcsname
  \csname #3\endcsname}%
\fi
\fi\endgroup
{\renewcommand{\dashlinestretch}{30}
\begin{picture}(13610,9694)(0,-10)
\put(10088,4062){\makebox(0,0)[b]{\smash{{{\SetFigFont{5}{6.0}{rm}$v$}}}}}
\put(5138,912){\blacken\ellipse{180}{180}}
\put(5138,912){\ellipse{180}{180}}
\put(6488,642){\blacken\ellipse{180}{180}}
\put(6488,642){\ellipse{180}{180}}
\put(7928,1362){\blacken\ellipse{180}{180}}
\put(7928,1362){\ellipse{180}{180}}
\put(7928,372){\blacken\ellipse{180}{180}}
\put(7928,372){\ellipse{180}{180}}
\put(6488,1272){\blacken\ellipse{180}{180}}
\put(6488,1272){\ellipse{180}{180}}
\put(6218,192){\arc{360}{1.5708}{3.1416}}
\put(6218,1632){\arc{360}{3.1416}{4.7124}}
\put(6848,1632){\arc{360}{4.7124}{6.2832}}
\put(6848,192){\arc{360}{0}{1.5708}}
\path(6038,192)(6038,1632)
\path(6218,1812)(6848,1812)
\path(7028,1632)(7028,192)
\path(6848,12)(6218,12)
\put(7838,192){\arc{360}{1.5708}{3.1416}}
\put(7838,1632){\arc{360}{3.1416}{4.7124}}
\put(8468,1632){\arc{360}{4.7124}{6.2832}}
\put(8468,192){\arc{360}{0}{1.5708}}
\path(7658,192)(7658,1632)
\path(7838,1812)(8468,1812)
\path(8648,1632)(8648,192)
\path(8468,12)(7838,12)
\put(6578,2262){\makebox(0,0)[b]{\smash{{{\SetFigFont{5}{6.0}{rm}$N(v)$}}}}}
\put(8108,2262){\makebox(0,0)[b]{\smash{{{\SetFigFont{5}{6.0}{rm}$\overline{N(v)}$}}}}}
\put(5138,462){\makebox(0,0)[b]{\smash{{{\SetFigFont{5}{6.0}{rm}$v$}}}}}
\put(13328,822){\blacken\ellipse{180}{180}}
\put(13328,822){\ellipse{180}{180}}
\put(9998,912){\blacken\ellipse{180}{180}}
\put(9998,912){\ellipse{180}{180}}
\put(11348,642){\blacken\ellipse{180}{180}}
\put(11348,642){\ellipse{180}{180}}
\put(12788,1362){\blacken\ellipse{180}{180}}
\put(12788,1362){\ellipse{180}{180}}
\put(12788,372){\blacken\ellipse{180}{180}}
\put(12788,372){\ellipse{180}{180}}
\put(11348,1272){\blacken\ellipse{180}{180}}
\put(11348,1272){\ellipse{180}{180}}
\put(11078,192){\arc{360}{1.5708}{3.1416}}
\put(11078,1632){\arc{360}{3.1416}{4.7124}}
\put(11708,1632){\arc{360}{4.7124}{6.2832}}
\put(11708,192){\arc{360}{0}{1.5708}}
\path(10898,192)(10898,1632)
\path(11078,1812)(11708,1812)
\path(11888,1632)(11888,192)
\path(11708,12)(11078,12)
\put(12698,192){\arc{360}{1.5708}{3.1416}}
\put(12698,1632){\arc{360}{3.1416}{4.7124}}
\put(13328,1632){\arc{360}{4.7124}{6.2832}}
\put(13328,192){\arc{360}{0}{1.5708}}
\path(12518,192)(12518,1632)
\path(12698,1812)(13328,1812)
\path(13508,1632)(13508,192)
\path(13328,12)(12698,12)
\put(11438,2262){\makebox(0,0)[b]{\smash{{{\SetFigFont{5}{6.0}{rm}$N(v)$}}}}}
\put(12968,2262){\makebox(0,0)[b]{\smash{{{\SetFigFont{5}{6.0}{rm}$\overline{N(v)}$}}}}}
\put(9998,462){\makebox(0,0)[b]{\smash{{{\SetFigFont{5}{6.0}{rm}$v$}}}}}
\put(1538,1272){\blacken\ellipse{180}{180}}
\put(1538,1272){\ellipse{180}{180}}
\put(2978,372){\blacken\ellipse{180}{180}}
\put(2978,372){\ellipse{180}{180}}
\put(2978,1362){\blacken\ellipse{180}{180}}
\put(2978,1362){\ellipse{180}{180}}
\put(1538,642){\blacken\ellipse{180}{180}}
\put(1538,642){\ellipse{180}{180}}
\put(188,912){\blacken\ellipse{180}{180}}
\put(188,912){\ellipse{180}{180}}
\put(3518,822){\blacken\ellipse{180}{180}}
\put(3518,822){\ellipse{180}{180}}
\put(2888,192){\arc{360}{1.5708}{3.1416}}
\put(2888,1632){\arc{360}{3.1416}{4.7124}}
\put(3518,1632){\arc{360}{4.7124}{6.2832}}
\put(3518,192){\arc{360}{0}{1.5708}}
\path(2708,192)(2708,1632)
\path(2888,1812)(3518,1812)
\path(3698,1632)(3698,192)
\path(3518,12)(2888,12)
\put(1268,192){\arc{360}{1.5708}{3.1416}}
\put(1268,1632){\arc{360}{3.1416}{4.7124}}
\put(1898,1632){\arc{360}{4.7124}{6.2832}}
\put(1898,192){\arc{360}{0}{1.5708}}
\path(1088,192)(1088,1632)
\path(1268,1812)(1898,1812)
\path(2078,1632)(2078,192)
\path(1898,12)(1268,12)
\put(188,462){\makebox(0,0)[b]{\smash{{{\SetFigFont{5}{6.0}{rm}$v$}}}}}
\put(3158,2262){\makebox(0,0)[b]{\smash{{{\SetFigFont{5}{6.0}{rm}$\overline{N(v)}$}}}}}
\put(1628,2262){\makebox(0,0)[b]{\smash{{{\SetFigFont{5}{6.0}{rm}$N(v)$}}}}}
\put(3338,8652){\blacken\ellipse{180}{180}}
\put(3338,8652){\ellipse{180}{180}}
\put(2888,7752){\blacken\ellipse{180}{180}}
\put(2888,7752){\ellipse{180}{180}}
\put(2888,8652){\blacken\ellipse{180}{180}}
\put(2888,8652){\ellipse{180}{180}}
\put(1448,8022){\blacken\ellipse{180}{180}}
\put(1448,8022){\ellipse{180}{180}}
\put(98,8022){\blacken\ellipse{180}{180}}
\put(98,8022){\ellipse{180}{180}}
\put(3338,7752){\blacken\ellipse{180}{180}}
\put(3338,7752){\ellipse{180}{180}}
\put(7838,7392){\blacken\ellipse{180}{180}}
\put(7838,7392){\ellipse{180}{180}}
\put(5048,8022){\blacken\ellipse{180}{180}}
\put(5048,8022){\ellipse{180}{180}}
\put(6398,8022){\blacken\ellipse{180}{180}}
\put(6398,8022){\ellipse{180}{180}}
\put(7838,8742){\blacken\ellipse{180}{180}}
\put(7838,8742){\ellipse{180}{180}}
\put(8288,8742){\blacken\ellipse{180}{180}}
\put(8288,8742){\ellipse{180}{180}}
\put(7838,8292){\blacken\ellipse{180}{180}}
\put(7838,8292){\ellipse{180}{180}}
\put(12878,7572){\blacken\ellipse{180}{180}}
\put(12878,7572){\ellipse{180}{180}}
\put(10088,8022){\blacken\ellipse{180}{180}}
\put(10088,8022){\ellipse{180}{180}}
\put(11528,8022){\blacken\ellipse{180}{180}}
\put(11528,8022){\ellipse{180}{180}}
\put(12878,8472){\blacken\ellipse{180}{180}}
\put(12878,8472){\ellipse{180}{180}}
\put(12878,8022){\blacken\ellipse{180}{180}}
\put(12878,8022){\ellipse{180}{180}}
\put(13418,8022){\blacken\ellipse{180}{180}}
\put(13418,8022){\ellipse{180}{180}}
\put(3518,4512){\blacken\ellipse{180}{180}}
\put(3518,4512){\ellipse{180}{180}}
\put(188,4602){\blacken\ellipse{180}{180}}
\put(188,4602){\ellipse{180}{180}}
\put(1538,4332){\blacken\ellipse{180}{180}}
\put(1538,4332){\ellipse{180}{180}}
\put(2978,5052){\blacken\ellipse{180}{180}}
\put(2978,5052){\ellipse{180}{180}}
\put(2978,4332){\blacken\ellipse{180}{180}}
\put(2978,4332){\ellipse{180}{180}}
\put(1538,4962){\blacken\ellipse{180}{180}}
\put(1538,4962){\ellipse{180}{180}}
\put(7838,4422){\blacken\ellipse{180}{180}}
\put(7838,4422){\ellipse{180}{180}}
\put(5048,4512){\blacken\ellipse{180}{180}}
\put(5048,4512){\ellipse{180}{180}}
\put(6398,4242){\blacken\ellipse{180}{180}}
\put(6398,4242){\ellipse{180}{180}}
\put(7838,4962){\blacken\ellipse{180}{180}}
\put(7838,4962){\ellipse{180}{180}}
\put(7838,3972){\blacken\ellipse{180}{180}}
\put(7838,3972){\ellipse{180}{180}}
\put(6398,4872){\blacken\ellipse{180}{180}}
\put(6398,4872){\ellipse{180}{180}}
\put(13418,4422){\blacken\ellipse{180}{180}}
\put(13418,4422){\ellipse{180}{180}}
\put(10088,4512){\blacken\ellipse{180}{180}}
\put(10088,4512){\ellipse{180}{180}}
\put(11438,4242){\blacken\ellipse{180}{180}}
\put(11438,4242){\ellipse{180}{180}}
\put(12878,4962){\blacken\ellipse{180}{180}}
\put(12878,4962){\ellipse{180}{180}}
\put(12878,3972){\blacken\ellipse{180}{180}}
\put(12878,3972){\ellipse{180}{180}}
\put(11438,4872){\blacken\ellipse{180}{180}}
\put(11438,4872){\ellipse{180}{180}}
\path(12788,1362)(12788,372)
\path(6488,642)(7928,1362)
\path(3518,822)(2978,1362)
\path(6398,4872)(7838,3972)
\path(7838,4422)(7838,4962)(6398,4872)
	(6398,4242)(7838,4962)
\path(2978,5052)(2978,4332)(1538,4962)
\path(1538,4332)(1538,4962)(2978,5052)(3608,4512)
\path(11438,4872)(11438,4242)
\path(12878,8472)(13418,8022)(12878,8022)
\path(10088,8022)(11528,8022)(12878,8472)
	(12878,8022)(12878,7572)
\path(6398,8022)(7838,7392)
\path(7838,8742)(7838,8292)(6488,8022)
\path(4958,8022)(6398,8022)(7838,8742)(8288,8742)
\path(2888,7752)(2888,8562)
\path(1448,8022)(2888,7752)(3248,7752)
\path(98,8022)(1448,8022)(2888,8652)(3338,8652)
\put(2798,7302){\arc{360}{1.5708}{3.1416}}
\put(2798,8742){\arc{360}{3.1416}{4.7124}}
\put(3428,8742){\arc{360}{4.7124}{6.2832}}
\put(3428,7302){\arc{360}{0}{1.5708}}
\path(2618,7302)(2618,8742)
\path(2798,8922)(3428,8922)
\path(3608,8742)(3608,7302)
\path(3428,7122)(2798,7122)
\put(1178,7302){\arc{360}{1.5708}{3.1416}}
\put(1178,8742){\arc{360}{3.1416}{4.7124}}
\put(1808,8742){\arc{360}{4.7124}{6.2832}}
\put(1808,7302){\arc{360}{0}{1.5708}}
\path(998,7302)(998,8742)
\path(1178,8922)(1808,8922)
\path(1988,8742)(1988,7302)
\path(1808,7122)(1178,7122)
\put(6128,7392){\arc{360}{1.5708}{3.1416}}
\put(6128,8832){\arc{360}{3.1416}{4.7124}}
\put(6758,8832){\arc{360}{4.7124}{6.2832}}
\put(6758,7392){\arc{360}{0}{1.5708}}
\path(5948,7392)(5948,8832)
\path(6128,9012)(6758,9012)
\path(6938,8832)(6938,7392)
\path(6758,7212)(6128,7212)
\put(7748,7392){\arc{360}{1.5708}{3.1416}}
\put(7748,8832){\arc{360}{3.1416}{4.7124}}
\put(8378,8832){\arc{360}{4.7124}{6.2832}}
\put(8378,7392){\arc{360}{0}{1.5708}}
\path(7568,7392)(7568,8832)
\path(7748,9012)(8378,9012)
\path(8558,8832)(8558,7392)
\path(8378,7212)(7748,7212)
\put(11168,7392){\arc{360}{1.5708}{3.1416}}
\put(11168,8832){\arc{360}{3.1416}{4.7124}}
\put(11798,8832){\arc{360}{4.7124}{6.2832}}
\put(11798,7392){\arc{360}{0}{1.5708}}
\path(10988,7392)(10988,8832)
\path(11168,9012)(11798,9012)
\path(11978,8832)(11978,7392)
\path(11798,7212)(11168,7212)
\put(12788,7392){\arc{360}{1.5708}{3.1416}}
\put(12788,8832){\arc{360}{3.1416}{4.7124}}
\put(13418,8832){\arc{360}{4.7124}{6.2832}}
\put(13418,7392){\arc{360}{0}{1.5708}}
\path(12608,7392)(12608,8832)
\path(12788,9012)(13418,9012)
\path(13598,8832)(13598,7392)
\path(13418,7212)(12788,7212)
\put(1268,3882){\arc{360}{1.5708}{3.1416}}
\put(1268,5322){\arc{360}{3.1416}{4.7124}}
\put(1898,5322){\arc{360}{4.7124}{6.2832}}
\put(1898,3882){\arc{360}{0}{1.5708}}
\path(1088,3882)(1088,5322)
\path(1268,5502)(1898,5502)
\path(2078,5322)(2078,3882)
\path(1898,3702)(1268,3702)
\put(2888,3882){\arc{360}{1.5708}{3.1416}}
\put(2888,5322){\arc{360}{3.1416}{4.7124}}
\put(3518,5322){\arc{360}{4.7124}{6.2832}}
\put(3518,3882){\arc{360}{0}{1.5708}}
\path(2708,3882)(2708,5322)
\path(2888,5502)(3518,5502)
\path(3698,5322)(3698,3882)
\path(3518,3702)(2888,3702)
\put(6128,3792){\arc{360}{1.5708}{3.1416}}
\put(6128,5232){\arc{360}{3.1416}{4.7124}}
\put(6758,5232){\arc{360}{4.7124}{6.2832}}
\put(6758,3792){\arc{360}{0}{1.5708}}
\path(5948,3792)(5948,5232)
\path(6128,5412)(6758,5412)
\path(6938,5232)(6938,3792)
\path(6758,3612)(6128,3612)
\put(7748,3792){\arc{360}{1.5708}{3.1416}}
\put(7748,5232){\arc{360}{3.1416}{4.7124}}
\put(8378,5232){\arc{360}{4.7124}{6.2832}}
\put(8378,3792){\arc{360}{0}{1.5708}}
\path(7568,3792)(7568,5232)
\path(7748,5412)(8378,5412)
\path(8558,5232)(8558,3792)
\path(8378,3612)(7748,3612)
\put(11168,3792){\arc{360}{1.5708}{3.1416}}
\put(11168,5232){\arc{360}{3.1416}{4.7124}}
\put(11798,5232){\arc{360}{4.7124}{6.2832}}
\put(11798,3792){\arc{360}{0}{1.5708}}
\path(10988,3792)(10988,5232)
\path(11168,5412)(11798,5412)
\path(11978,5232)(11978,3792)
\path(11798,3612)(11168,3612)
\put(12788,3792){\arc{360}{1.5708}{3.1416}}
\put(12788,5232){\arc{360}{3.1416}{4.7124}}
\put(13418,5232){\arc{360}{4.7124}{6.2832}}
\put(13418,3792){\arc{360}{0}{1.5708}}
\path(12608,3792)(12608,5232)
\path(12788,5412)(13418,5412)
\path(13598,5232)(13598,3792)
\path(13418,3612)(12788,3612)
\path(1538,4962)(188,4602)(1538,4332)
\path(6398,4872)(5048,4512)(6398,4242)
\path(12878,4962)(11438,4872)(13418,4422)
	(11438,4242)(12878,3972)
\path(11438,4872)(10088,4512)(11438,4242)
\path(2978,372)(3518,822)(1538,642)
	(3068,1362)(1538,1272)(188,912)(1538,642)
\path(7928,372)(6488,642)(5138,912)
	(6398,1272)(7928,1362)(8468,822)(6488,642)
\path(9998,912)(11348,1272)(12788,1362)
	(13328,822)(12788,372)(11348,642)(9998,912)
\put(98,7572){\makebox(0,0)[b]{\smash{{{\SetFigFont{5}{6.0}{rm}$v$}}}}}
\put(1538,9372){\makebox(0,0)[b]{\smash{{{\SetFigFont{5}{6.0}{rm}$N(v)$}}}}}
\put(3068,9372){\makebox(0,0)[b]{\smash{{{\SetFigFont{5}{6.0}{rm}$\overline{N(v)}$}}}}}
\put(6488,9462){\makebox(0,0)[b]{\smash{{{\SetFigFont{5}{6.0}{rm}$N(v)$}}}}}
\put(5048,7662){\makebox(0,0)[b]{\smash{{{\SetFigFont{5}{6.0}{rm}$v$}}}}}
\put(11528,9462){\makebox(0,0)[b]{\smash{{{\SetFigFont{5}{6.0}{rm}$N(v)$}}}}}
\put(10088,7662){\makebox(0,0)[b]{\smash{{{\SetFigFont{5}{6.0}{rm}$v$}}}}}
\put(8018,9372){\makebox(0,0)[b]{\smash{{{\SetFigFont{5}{6.0}{rm}$\overline{N(v)}$}}}}}
\put(13058,9372){\makebox(0,0)[b]{\smash{{{\SetFigFont{5}{6.0}{rm}$\overline{N(v)}$}}}}}
\put(1628,5952){\makebox(0,0)[b]{\smash{{{\SetFigFont{5}{6.0}{rm}$N(v)$}}}}}
\put(3158,5952){\makebox(0,0)[b]{\smash{{{\SetFigFont{5}{6.0}{rm}$\overline{N(v)}$}}}}}
\put(188,4152){\makebox(0,0)[b]{\smash{{{\SetFigFont{5}{6.0}{rm}$v$}}}}}
\put(6488,5862){\makebox(0,0)[b]{\smash{{{\SetFigFont{5}{6.0}{rm}$N(v)$}}}}}
\put(8018,5862){\makebox(0,0)[b]{\smash{{{\SetFigFont{5}{6.0}{rm}$\overline{N(v)}$}}}}}
\put(5048,4062){\makebox(0,0)[b]{\smash{{{\SetFigFont{5}{6.0}{rm}$v$}}}}}
\put(11528,5862){\makebox(0,0)[b]{\smash{{{\SetFigFont{5}{6.0}{rm}$N(v)$}}}}}
\put(13058,5862){\makebox(0,0)[b]{\smash{{{\SetFigFont{5}{6.0}{rm}$\overline{N(v)}$}}}}}
\put(8468,822){\blacken\ellipse{180}{180}}
\put(8468,822){\ellipse{180}{180}}
\end{picture}
}

%% file: SommetvPasSurHouse.eepic
\setlength{\unitlength}{0.00026247in}
\begingroup\makeatletter\ifx\SetFigFont\undefined
\def\x#1#2#3#4#5#6#7\relax{\def\x{#1#2#3#4#5#6}}%
\expandafter\x\fmtname xxxxxx\relax \def\y{splain}%
\ifx\x\y   
\gdef\SetFigFont#1#2#3{%
  \ifnum #1<17\tiny\else \ifnum #1<20\small\else
  \ifnum #1<24\normalsize\else \ifnum #1<29\large\else
  \ifnum #1<34\Large\else \ifnum #1<41\LARGE\else
     \huge\fi\fi\fi\fi\fi\fi
  \csname #3\endcsname}%
\else
\gdef\SetFigFont#1#2#3{\begingroup
  \count@#1\relax \ifnum 25<\count@\count@25\fi
  \def\x{\endgroup\@setsize\SetFigFont{#2pt}}%
  \expandafter\x
    \csname \romannumeral\the\count@ pt\expandafter\endcsname
    \csname @\romannumeral\the\count@ pt\endcsname
  \csname #3\endcsname}%
\fi
\fi\endgroup
{\renewcommand{\dashlinestretch}{30}
\begin{picture}(13610,9694)(0,-10)
\put(10088,4062){\makebox(0,0)[b]{\smash{{{\SetFigFont{5}{6.0}{rm}$v$}}}}}
\put(8288,282){\blacken\ellipse{180}{180}}
\put(8288,282){\ellipse{180}{180}}
\put(8198,1452){\blacken\ellipse{180}{180}}
\put(8198,1452){\ellipse{180}{180}}
\put(6488,642){\blacken\ellipse{180}{180}}
\put(6488,642){\ellipse{180}{180}}
\put(5138,912){\blacken\ellipse{180}{180}}
\put(5138,912){\ellipse{180}{180}}
\put(7928,822){\blacken\ellipse{180}{180}}
\put(7928,822){\ellipse{180}{180}}
\put(3428,912){\blacken\ellipse{180}{180}}
\put(3428,912){\ellipse{180}{180}}
\put(188,912){\blacken\ellipse{180}{180}}
\put(188,912){\ellipse{180}{180}}
\put(1538,642){\blacken\ellipse{180}{180}}
\put(1538,642){\ellipse{180}{180}}
\put(2978,1362){\blacken\ellipse{180}{180}}
\put(2978,1362){\ellipse{180}{180}}
\put(2978,372){\blacken\ellipse{180}{180}}
\put(2978,372){\ellipse{180}{180}}
\put(1538,1272){\blacken\ellipse{180}{180}}
\put(1538,1272){\ellipse{180}{180}}
\put(3338,8652){\blacken\ellipse{180}{180}}
\put(3338,8652){\ellipse{180}{180}}
\put(2888,7752){\blacken\ellipse{180}{180}}
\put(2888,7752){\ellipse{180}{180}}
\put(2888,8652){\blacken\ellipse{180}{180}}
\put(2888,8652){\ellipse{180}{180}}
\put(1448,8022){\blacken\ellipse{180}{180}}
\put(1448,8022){\ellipse{180}{180}}
\put(98,8022){\blacken\ellipse{180}{180}}
\put(98,8022){\ellipse{180}{180}}
\put(3338,7752){\blacken\ellipse{180}{180}}
\put(3338,7752){\ellipse{180}{180}}
\put(7838,7392){\blacken\ellipse{180}{180}}
\put(7838,7392){\ellipse{180}{180}}
\put(5048,8022){\blacken\ellipse{180}{180}}
\put(5048,8022){\ellipse{180}{180}}
\put(6398,8022){\blacken\ellipse{180}{180}}
\put(6398,8022){\ellipse{180}{180}}
\put(7838,8742){\blacken\ellipse{180}{180}}
\put(7838,8742){\ellipse{180}{180}}
\put(8288,8742){\blacken\ellipse{180}{180}}
\put(8288,8742){\ellipse{180}{180}}
\put(7838,8292){\blacken\ellipse{180}{180}}
\put(7838,8292){\ellipse{180}{180}}
\put(13238,7482){\blacken\ellipse{180}{180}}
\put(13238,7482){\ellipse{180}{180}}
\put(10088,8022){\blacken\ellipse{180}{180}}
\put(10088,8022){\ellipse{180}{180}}
\put(11528,8022){\blacken\ellipse{180}{180}}
\put(11528,8022){\ellipse{180}{180}}
\put(12878,8472){\blacken\ellipse{180}{180}}
\put(12878,8472){\ellipse{180}{180}}
\put(12878,7842){\blacken\ellipse{180}{180}}
\put(12878,7842){\ellipse{180}{180}}
\put(13418,8292){\blacken\ellipse{180}{180}}
\put(13418,8292){\ellipse{180}{180}}
\put(2978,4512){\blacken\ellipse{180}{180}}
\put(2978,4512){\ellipse{180}{180}}
\put(188,4602){\blacken\ellipse{180}{180}}
\put(188,4602){\ellipse{180}{180}}
\put(1538,4332){\blacken\ellipse{180}{180}}
\put(1538,4332){\ellipse{180}{180}}
\put(3428,5052){\blacken\ellipse{180}{180}}
\put(3428,5052){\ellipse{180}{180}}
\put(3428,3972){\blacken\ellipse{180}{180}}
\put(3428,3972){\ellipse{180}{180}}
\put(1538,4962){\blacken\ellipse{180}{180}}
\put(1538,4962){\ellipse{180}{180}}
\put(7838,4422){\blacken\ellipse{180}{180}}
\put(7838,4422){\ellipse{180}{180}}
\put(5048,4512){\blacken\ellipse{180}{180}}
\put(5048,4512){\ellipse{180}{180}}
\put(6398,4242){\blacken\ellipse{180}{180}}
\put(6398,4242){\ellipse{180}{180}}
\put(7838,4962){\blacken\ellipse{180}{180}}
\put(7838,4962){\ellipse{180}{180}}
\put(7838,3972){\blacken\ellipse{180}{180}}
\put(7838,3972){\ellipse{180}{180}}
\put(6398,4872){\blacken\ellipse{180}{180}}
\put(6398,4872){\ellipse{180}{180}}
\put(13328,4512){\blacken\ellipse{180}{180}}
\put(13328,4512){\ellipse{180}{180}}
\put(10088,4512){\blacken\ellipse{180}{180}}
\put(10088,4512){\ellipse{180}{180}}
\put(11438,4242){\blacken\ellipse{180}{180}}
\put(11438,4242){\ellipse{180}{180}}
\put(12878,4872){\blacken\ellipse{180}{180}}
\put(12878,4872){\ellipse{180}{180}}
\put(12878,4152){\blacken\ellipse{180}{180}}
\put(12878,4152){\ellipse{180}{180}}
\put(11438,4872){\blacken\ellipse{180}{180}}
\put(11438,4872){\ellipse{180}{180}}
\path(6488,1272)(7928,822)(6488,642)
\path(6488,642)(8198,1452)
\path(5138,912)(6488,1272)(8198,1452)
	(8288,192)(6488,642)(5138,912)
\put(7838,192){\arc{360}{1.5708}{3.1416}}
\put(7838,1632){\arc{360}{3.1416}{4.7124}}
\put(8468,1632){\arc{360}{4.7124}{6.2832}}
\put(8468,192){\arc{360}{0}{1.5708}}
\path(7658,192)(7658,1632)
\path(7838,1812)(8468,1812)
\path(8648,1632)(8648,192)
\path(8468,12)(7838,12)
\put(6218,192){\arc{360}{1.5708}{3.1416}}
\put(6218,1632){\arc{360}{3.1416}{4.7124}}
\put(6848,1632){\arc{360}{4.7124}{6.2832}}
\put(6848,192){\arc{360}{0}{1.5708}}
\path(6038,192)(6038,1632)
\path(6218,1812)(6848,1812)
\path(7028,1632)(7028,192)
\path(6848,12)(6218,12)
\path(1538,642)(2978,1362)
\path(1538,1272)(3428,912)
\path(1538,642)(188,912)(1538,1272)
	(2978,1362)(3428,912)(2978,372)(1538,642)
\put(1268,192){\arc{360}{1.5708}{3.1416}}
\put(1268,1632){\arc{360}{3.1416}{4.7124}}
\put(1898,1632){\arc{360}{4.7124}{6.2832}}
\put(1898,192){\arc{360}{0}{1.5708}}
\path(1088,192)(1088,1632)
\path(1268,1812)(1898,1812)
\path(2078,1632)(2078,192)
\path(1898,12)(1268,12)
\put(2888,192){\arc{360}{1.5708}{3.1416}}
\put(2888,1632){\arc{360}{3.1416}{4.7124}}
\put(3518,1632){\arc{360}{4.7124}{6.2832}}
\put(3518,192){\arc{360}{0}{1.5708}}
\path(2708,192)(2708,1632)
\path(2888,1812)(3518,1812)
\path(3698,1632)(3698,192)
\path(3518,12)(2888,12)
\path(12878,4152)(11528,4242)
\path(11438,4872)(12878,4872)(12878,4152)
	(13328,4512)(12878,4872)
\path(7838,4422)(7838,4062)
\path(1538,4332)(2978,4512)(1538,4962)(1538,4332)
\path(3428,3972)(3428,5052)(1628,4962)
	(188,4602)(1538,4332)(3428,3972)
\path(13418,8292)(12878,7842)
\path(10088,8022)(11528,8022)(12878,8472)
	(13418,8292)(13238,7482)(12878,7842)(11528,8022)
\path(8288,8652)(7838,7392)
\path(3338,8652)(3338,7752)
\path(6398,4872)(7838,3972)
\path(7838,4422)(7838,4962)(6398,4872)
	(6398,4242)(7838,4962)
\path(11438,4872)(11438,4242)
\path(6398,8022)(7838,7392)
\path(7838,8742)(7838,8292)(6488,8022)
\path(4958,8022)(6398,8022)(7838,8742)(8288,8742)
\path(2888,7752)(2888,8562)
\path(1448,8022)(2888,7752)(3248,7752)
\path(98,8022)(1448,8022)(2888,8652)(3338,8652)
\put(2798,7302){\arc{360}{1.5708}{3.1416}}
\put(2798,8742){\arc{360}{3.1416}{4.7124}}
\put(3428,8742){\arc{360}{4.7124}{6.2832}}
\put(3428,7302){\arc{360}{0}{1.5708}}
\path(2618,7302)(2618,8742)
\path(2798,8922)(3428,8922)
\path(3608,8742)(3608,7302)
\path(3428,7122)(2798,7122)
\put(1178,7302){\arc{360}{1.5708}{3.1416}}
\put(1178,8742){\arc{360}{3.1416}{4.7124}}
\put(1808,8742){\arc{360}{4.7124}{6.2832}}
\put(1808,7302){\arc{360}{0}{1.5708}}
\path(998,7302)(998,8742)
\path(1178,8922)(1808,8922)
\path(1988,8742)(1988,7302)
\path(1808,7122)(1178,7122)
\put(6128,7392){\arc{360}{1.5708}{3.1416}}
\put(6128,8832){\arc{360}{3.1416}{4.7124}}
\put(6758,8832){\arc{360}{4.7124}{6.2832}}
\put(6758,7392){\arc{360}{0}{1.5708}}
\path(5948,7392)(5948,8832)
\path(6128,9012)(6758,9012)
\path(6938,8832)(6938,7392)
\path(6758,7212)(6128,7212)
\put(7748,7392){\arc{360}{1.5708}{3.1416}}
\put(7748,8832){\arc{360}{3.1416}{4.7124}}
\put(8378,8832){\arc{360}{4.7124}{6.2832}}
\put(8378,7392){\arc{360}{0}{1.5708}}
\path(7568,7392)(7568,8832)
\path(7748,9012)(8378,9012)
\path(8558,8832)(8558,7392)
\path(8378,7212)(7748,7212)
\put(11168,7392){\arc{360}{1.5708}{3.1416}}
\put(11168,8832){\arc{360}{3.1416}{4.7124}}
\put(11798,8832){\arc{360}{4.7124}{6.2832}}
\put(11798,7392){\arc{360}{0}{1.5708}}
\path(10988,7392)(10988,8832)
\path(11168,9012)(11798,9012)
\path(11978,8832)(11978,7392)
\path(11798,7212)(11168,7212)
\put(12788,7392){\arc{360}{1.5708}{3.1416}}
\put(12788,8832){\arc{360}{3.1416}{4.7124}}
\put(13418,8832){\arc{360}{4.7124}{6.2832}}
\put(13418,7392){\arc{360}{0}{1.5708}}
\path(12608,7392)(12608,8832)
\path(12788,9012)(13418,9012)
\path(13598,8832)(13598,7392)
\path(13418,7212)(12788,7212)
\put(1268,3882){\arc{360}{1.5708}{3.1416}}
\put(1268,5322){\arc{360}{3.1416}{4.7124}}
\put(1898,5322){\arc{360}{4.7124}{6.2832}}
\put(1898,3882){\arc{360}{0}{1.5708}}
\path(1088,3882)(1088,5322)
\path(1268,5502)(1898,5502)
\path(2078,5322)(2078,3882)
\path(1898,3702)(1268,3702)
\put(2888,3882){\arc{360}{1.5708}{3.1416}}
\put(2888,5322){\arc{360}{3.1416}{4.7124}}
\put(3518,5322){\arc{360}{4.7124}{6.2832}}
\put(3518,3882){\arc{360}{0}{1.5708}}
\path(2708,3882)(2708,5322)
\path(2888,5502)(3518,5502)
\path(3698,5322)(3698,3882)
\path(3518,3702)(2888,3702)
\put(6128,3792){\arc{360}{1.5708}{3.1416}}
\put(6128,5232){\arc{360}{3.1416}{4.7124}}
\put(6758,5232){\arc{360}{4.7124}{6.2832}}
\put(6758,3792){\arc{360}{0}{1.5708}}
\path(5948,3792)(5948,5232)
\path(6128,5412)(6758,5412)
\path(6938,5232)(6938,3792)
\path(6758,3612)(6128,3612)
\put(7748,3792){\arc{360}{1.5708}{3.1416}}
\put(7748,5232){\arc{360}{3.1416}{4.7124}}
\put(8378,5232){\arc{360}{4.7124}{6.2832}}
\put(8378,3792){\arc{360}{0}{1.5708}}
\path(7568,3792)(7568,5232)
\path(7748,5412)(8378,5412)
\path(8558,5232)(8558,3792)
\path(8378,3612)(7748,3612)
\put(11168,3792){\arc{360}{1.5708}{3.1416}}
\put(11168,5232){\arc{360}{3.1416}{4.7124}}
\put(11798,5232){\arc{360}{4.7124}{6.2832}}
\put(11798,3792){\arc{360}{0}{1.5708}}
\path(10988,3792)(10988,5232)
\path(11168,5412)(11798,5412)
\path(11978,5232)(11978,3792)
\path(11798,3612)(11168,3612)
\put(12788,3792){\arc{360}{1.5708}{3.1416}}
\put(12788,5232){\arc{360}{3.1416}{4.7124}}
\put(13418,5232){\arc{360}{4.7124}{6.2832}}
\put(13418,3792){\arc{360}{0}{1.5708}}
\path(12608,3792)(12608,5232)
\path(12788,5412)(13418,5412)
\path(13598,5232)(13598,3792)
\path(13418,3612)(12788,3612)
\path(6398,4872)(5048,4512)(6398,4242)
\path(11438,4872)(10088,4512)(11438,4242)
\put(5138,462){\makebox(0,0)[b]{\smash{{{\SetFigFont{5}{6.0}{rm}$v$}}}}}
\put(8108,2262){\makebox(0,0)[b]{\smash{{{\SetFigFont{5}{6.0}{rm}$\overline{N(v)}$}}}}}
\put(6578,2262){\makebox(0,0)[b]{\smash{{{\SetFigFont{5}{6.0}{rm}$N(v)$}}}}}
\put(1628,2262){\makebox(0,0)[b]{\smash{{{\SetFigFont{5}{6.0}{rm}$N(v)$}}}}}
\put(3158,2262){\makebox(0,0)[b]{\smash{{{\SetFigFont{5}{6.0}{rm}$\overline{N(v)}$}}}}}
\put(188,462){\makebox(0,0)[b]{\smash{{{\SetFigFont{5}{6.0}{rm}$v$}}}}}
\put(98,7572){\makebox(0,0)[b]{\smash{{{\SetFigFont{5}{6.0}{rm}$v$}}}}}
\put(1538,9372){\makebox(0,0)[b]{\smash{{{\SetFigFont{5}{6.0}{rm}$N(v)$}}}}}
\put(3068,9372){\makebox(0,0)[b]{\smash{{{\SetFigFont{5}{6.0}{rm}$\overline{N(v)}$}}}}}
\put(6488,9462){\makebox(0,0)[b]{\smash{{{\SetFigFont{5}{6.0}{rm}$N(v)$}}}}}
\put(5048,7662){\makebox(0,0)[b]{\smash{{{\SetFigFont{5}{6.0}{rm}$v$}}}}}
\put(11528,9462){\makebox(0,0)[b]{\smash{{{\SetFigFont{5}{6.0}{rm}$N(v)$}}}}}
\put(10088,7662){\makebox(0,0)[b]{\smash{{{\SetFigFont{5}{6.0}{rm}$v$}}}}}
\put(8018,9372){\makebox(0,0)[b]{\smash{{{\SetFigFont{5}{6.0}{rm}$\overline{N(v)}$}}}}}
\put(13058,9372){\makebox(0,0)[b]{\smash{{{\SetFigFont{5}{6.0}{rm}$\overline{N(v)}$}}}}}
\put(1628,5952){\makebox(0,0)[b]{\smash{{{\SetFigFont{5}{6.0}{rm}$N(v)$}}}}}
\put(3158,5952){\makebox(0,0)[b]{\smash{{{\SetFigFont{5}{6.0}{rm}$\overline{N(v)}$}}}}}
\put(188,4152){\makebox(0,0)[b]{\smash{{{\SetFigFont{5}{6.0}{rm}$v$}}}}}
\put(6488,5862){\makebox(0,0)[b]{\smash{{{\SetFigFont{5}{6.0}{rm}$N(v)$}}}}}
\put(8018,5862){\makebox(0,0)[b]{\smash{{{\SetFigFont{5}{6.0}{rm}$\overline{N(v)}$}}}}}
\put(5048,4062){\makebox(0,0)[b]{\smash{{{\SetFigFont{5}{6.0}{rm}$v$}}}}}
\put(11528,5862){\makebox(0,0)[b]{\smash{{{\SetFigFont{5}{6.0}{rm}$N(v)$}}}}}
\put(13058,5862){\makebox(0,0)[b]{\smash{{{\SetFigFont{5}{6.0}{rm}$\overline{N(v)}$}}}}}
\put(6488,1272){\blacken\ellipse{180}{180}}
\put(6488,1272){\ellipse{180}{180}}
\end{picture}
}

%% file: P5-freeBiparti.eepic
\setlength{\unitlength}{0.00026247in}
\begingroup\makeatletter\ifx\SetFigFont\undefined
\def\x#1#2#3#4#5#6#7\relax{\def\x{#1#2#3#4#5#6}}%
\expandafter\x\fmtname xxxxxx\relax \def\y{splain}%
\ifx\x\y   
\gdef\SetFigFont#1#2#3{%
  \ifnum #1<17\tiny\else \ifnum #1<20\small\else
  \ifnum #1<24\normalsize\else \ifnum #1<29\large\else
  \ifnum #1<34\Large\else \ifnum #1<41\LARGE\else
     \huge\fi\fi\fi\fi\fi\fi
  \csname #3\endcsname}%
\else
\gdef\SetFigFont#1#2#3{\begingroup
  \count@#1\relax \ifnum 25<\count@\count@25\fi
  \def\x{\endgroup\@setsize\SetFigFont{#2pt}}%
  \expandafter\x
    \csname \romannumeral\the\count@ pt\expandafter\endcsname
    \csname @\romannumeral\the\count@ pt\endcsname
  \csname #3\endcsname}%
\fi
\fi\endgroup
{\renewcommand{\dashlinestretch}{30}
\begin{picture}(2430,3925)(0,-10)
\put(0,3697){\makebox(0,0)[b]{\smash{{{\SetFigFont{5}{6.0}{rm}$b_1$}}}}}
\put(1890,97){\ellipse{180}{180}}
\put(540,97){\blacken\ellipse{180}{180}}
\put(540,97){\ellipse{180}{180}}
\put(540,3697){\blacken\ellipse{180}{180}}
\put(540,3697){\ellipse{180}{180}}
\put(540,2797){\blacken\ellipse{180}{180}}
\put(540,2797){\ellipse{180}{180}}
\put(540,1897){\blacken\ellipse{180}{180}}
\put(540,1897){\ellipse{180}{180}}
\put(540,997){\blacken\ellipse{180}{180}}
\put(540,997){\ellipse{180}{180}}
\put(1890,2797){\ellipse{180}{180}}
\put(1890,1897){\ellipse{180}{180}}
\put(1890,997){\ellipse{180}{180}}
\path(540,97)(1800,3697)
\path(540,97)(1800,2797)
\path(540,97)(1800,1897)
\path(540,97)(1800,997)
\path(1800,97)(540,97)
\path(1800,3697)(540,3697)
\path(1800,2797)(540,2797)
\path(1800,1897)(540,1897)
\path(1800,997)(540,997)
\path(1800,3697)(585,2797)
\path(1800,3697)(540,1897)
\path(1800,3697)(540,997)
\path(1800,2797)(540,1897)
\path(1800,2797)(540,952)
\path(1800,1897)(585,997)
\put(2430,3697){\makebox(0,0)[b]{\smash{{{\SetFigFont{5}{6.0}{rm}$w_5$}}}}}
\put(2430,2797){\makebox(0,0)[b]{\smash{{{\SetFigFont{5}{6.0}{rm}$w_4$}}}}}
\put(2430,1897){\makebox(0,0)[b]{\smash{{{\SetFigFont{5}{6.0}{rm}$w_3$}}}}}
\put(2430,997){\makebox(0,0)[b]{\smash{{{\SetFigFont{5}{6.0}{rm}$w_2$}}}}}
\put(0,97){\makebox(0,0)[b]{\smash{{{\SetFigFont{5}{6.0}{rm}$b_5$}}}}}
\put(0,997){\makebox(0,0)[b]{\smash{{{\SetFigFont{5}{6.0}{rm}$b_4$}}}}}
\put(0,1897){\makebox(0,0)[b]{\smash{{{\SetFigFont{5}{6.0}{rm}$b_3$}}}}}
\put(0,2797){\makebox(0,0)[b]{\smash{{{\SetFigFont{5}{6.0}{rm}$b_2$}}}}}
\put(2430,97){\makebox(0,0)[b]{\smash{{{\SetFigFont{5}{6.0}{rm}$w_1$}}}}}
\put(1890,3697){\ellipse{180}{180}}
\end{picture}
}

%% file: SeidelComplementationOnb2.eepic
\setlength{\unitlength}{0.00026247in}
\begingroup\makeatletter\ifx\SetFigFont\undefined
\def\x#1#2#3#4#5#6#7\relax{\def\x{#1#2#3#4#5#6}}%
\expandafter\x\fmtname xxxxxx\relax \def\y{splain}%
\ifx\x\y   
\gdef\SetFigFont#1#2#3{%
  \ifnum #1<17\tiny\else \ifnum #1<20\small\else
  \ifnum #1<24\normalsize\else \ifnum #1<29\large\else
  \ifnum #1<34\Large\else \ifnum #1<41\LARGE\else
     \huge\fi\fi\fi\fi\fi\fi
  \csname #3\endcsname}%
\else
\gdef\SetFigFont#1#2#3{\begingroup
  \count@#1\relax \ifnum 25<\count@\count@25\fi
  \def\x{\endgroup\@setsize\SetFigFont{#2pt}}%
  \expandafter\x
    \csname \romannumeral\the\count@ pt\expandafter\endcsname
    \csname @\romannumeral\the\count@ pt\endcsname
  \csname #3\endcsname}%
\fi
\fi\endgroup
{\renewcommand{\dashlinestretch}{30}
\begin{picture}(19440,8135)(0,-10)
\put(13005,2355){\makebox(0,0)[b]{\smash{{{\SetFigFont{5}{6.0}{rm}$w_2$}}}}}
\put(18810,3390){\blacken\ellipse{180}{180}}
\put(18810,3390){\ellipse{180}{180}}
\put(17550,2490){\ellipse{180}{180}}
\put(17550,3390){\ellipse{180}{180}}
\put(17550,3975){\ellipse{180}{180}}
\put(18765,3975){\blacken\ellipse{180}{180}}
\put(18765,3975){\ellipse{180}{180}}
\put(17550,5685){\blacken\ellipse{180}{180}}
\put(17550,5685){\ellipse{180}{180}}
\put(1800,6450){\ellipse{180}{180}}
\put(540,6450){\blacken\ellipse{180}{180}}
\put(540,6450){\ellipse{180}{180}}
\put(540,7350){\blacken\ellipse{180}{180}}
\put(540,7350){\ellipse{180}{180}}
\put(1800,7350){\ellipse{180}{180}}
\put(540,5235){\blacken\ellipse{180}{180}}
\put(540,5235){\ellipse{180}{180}}
\put(1755,5235){\ellipse{180}{180}}
\put(517,6000){\blacken\ellipse{44}{44}}
\put(517,6000){\ellipse{44}{44}}
\put(540,5685){\blacken\ellipse{44}{44}}
\put(540,5685){\ellipse{44}{44}}
\put(1755,6023){\blacken\ellipse{44}{44}}
\put(1755,6023){\ellipse{44}{44}}
\put(1732,5685){\blacken\ellipse{44}{44}}
\put(1732,5685){\ellipse{44}{44}}
\put(517,4673){\blacken\ellipse{44}{44}}
\put(517,4673){\ellipse{44}{44}}
\put(1755,4673){\blacken\ellipse{44}{44}}
\put(1755,4673){\ellipse{44}{44}}
\put(1800,3345){\ellipse{180}{180}}
\put(540,3345){\blacken\ellipse{180}{180}}
\put(540,3345){\ellipse{180}{180}}
\put(540,2445){\blacken\ellipse{180}{180}}
\put(540,2445){\ellipse{180}{180}}
\put(1800,2445){\ellipse{180}{180}}
\put(540,4110){\blacken\ellipse{44}{44}}
\put(540,4110){\ellipse{44}{44}}
\put(1777,4110){\blacken\ellipse{44}{44}}
\put(1777,4110){\ellipse{44}{44}}
\put(18810,6495){\ellipse{180}{180}}
\put(17550,6495){\blacken\ellipse{180}{180}}
\put(17550,6495){\ellipse{180}{180}}
\put(17550,7395){\blacken\ellipse{180}{180}}
\put(17550,7395){\ellipse{180}{180}}
\put(18810,7395){\ellipse{180}{180}}
\put(17550,5280){\blacken\ellipse{180}{180}}
\put(17550,5280){\ellipse{180}{180}}
\put(18765,5280){\ellipse{180}{180}}
\put(17527,6045){\blacken\ellipse{44}{44}}
\put(17527,6045){\ellipse{44}{44}}
\put(18765,6068){\blacken\ellipse{44}{44}}
\put(18765,6068){\ellipse{44}{44}}
\put(17527,4718){\blacken\ellipse{44}{44}}
\put(17527,4718){\ellipse{44}{44}}
\put(18765,4718){\blacken\ellipse{44}{44}}
\put(18765,4718){\ellipse{44}{44}}
\put(13680,5685){\blacken\ellipse{180}{180}}
\put(13680,5685){\ellipse{180}{180}}
\put(13680,6023){\blacken\ellipse{44}{44}}
\put(13680,6023){\ellipse{44}{44}}
\put(12285,6450){\ellipse{180}{180}}
\put(13680,6450){\blacken\ellipse{180}{180}}
\put(13680,6450){\ellipse{180}{180}}
\put(13635,7350){\blacken\ellipse{180}{180}}
\put(13635,7350){\ellipse{180}{180}}
\put(12285,7350){\ellipse{180}{180}}
\put(11025,5235){\blacken\ellipse{180}{180}}
\put(11025,5235){\ellipse{180}{180}}
\put(12240,5235){\ellipse{180}{180}}
\put(12240,6023){\blacken\ellipse{44}{44}}
\put(12240,6023){\ellipse{44}{44}}
\put(12217,5685){\blacken\ellipse{44}{44}}
\put(12217,5685){\ellipse{44}{44}}
\put(5512,5685){\blacken\ellipse{44}{44}}
\put(5512,5685){\ellipse{44}{44}}
\put(5535,6023){\blacken\ellipse{44}{44}}
\put(5535,6023){\ellipse{44}{44}}
\put(4320,5235){\blacken\ellipse{180}{180}}
\put(4320,5235){\ellipse{180}{180}}
\put(5580,7350){\ellipse{180}{180}}
\put(6930,7350){\blacken\ellipse{180}{180}}
\put(6930,7350){\ellipse{180}{180}}
\put(6975,6450){\blacken\ellipse{180}{180}}
\put(6975,6450){\ellipse{180}{180}}
\put(5580,6450){\ellipse{180}{180}}
\put(6975,6023){\blacken\ellipse{44}{44}}
\put(6975,6023){\ellipse{44}{44}}
\put(6975,5685){\blacken\ellipse{180}{180}}
\put(6975,5685){\ellipse{180}{180}}
\put(8235,4335){\ellipse{180}{180}}
\put(7020,4335){\blacken\ellipse{180}{180}}
\put(7020,4335){\ellipse{180}{180}}
\put(8257,3210){\blacken\ellipse{44}{44}}
\put(8257,3210){\ellipse{44}{44}}
\put(7020,3210){\blacken\ellipse{44}{44}}
\put(7020,3210){\ellipse{44}{44}}
\put(8280,1545){\ellipse{180}{180}}
\put(7020,1545){\blacken\ellipse{180}{180}}
\put(7020,1545){\ellipse{180}{180}}
\put(7020,2445){\blacken\ellipse{180}{180}}
\put(7020,2445){\ellipse{180}{180}}
\put(8280,2445){\ellipse{180}{180}}
\put(6997,3840){\blacken\ellipse{44}{44}}
\put(6997,3840){\ellipse{44}{44}}
\put(8212,3885){\blacken\ellipse{44}{44}}
\put(8212,3885){\ellipse{44}{44}}
\put(5535,5235){\ellipse{180}{180}}
\put(13702,3885){\blacken\ellipse{44}{44}}
\put(13702,3885){\ellipse{44}{44}}
\put(14917,3840){\blacken\ellipse{44}{44}}
\put(14917,3840){\ellipse{44}{44}}
\put(13635,2445){\ellipse{180}{180}}
\put(14895,2445){\blacken\ellipse{180}{180}}
\put(14895,2445){\ellipse{180}{180}}
\put(14895,1545){\blacken\ellipse{180}{180}}
\put(14895,1545){\ellipse{180}{180}}
\put(13635,1545){\ellipse{180}{180}}
\put(14895,3210){\blacken\ellipse{44}{44}}
\put(14895,3210){\ellipse{44}{44}}
\put(13657,3210){\blacken\ellipse{44}{44}}
\put(13657,3210){\ellipse{44}{44}}
\put(14895,4335){\blacken\ellipse{180}{180}}
\put(14895,4335){\ellipse{180}{180}}
\put(13680,4335){\ellipse{180}{180}}
\path(17550,3390)(18810,3975)
\path(17550,3975)(18810,3975)
\path(17550,2490)(18765,3975)
\path(1800,7350)(585,6450)
\path(1800,6450)(540,6450)
\path(1800,7350)(540,7350)
\path(1755,5235)(495,5235)
\path(517,5235)(1800,6450)
\path(517,5258)(1800,7350)
\path(1755,5235)(585,3345)
\path(1800,3345)(540,3345)
\path(1800,2445)(540,2445)
\path(540,2445)(1800,3345)
\path(540,2445)(1755,5235)
\path(18810,7395)(17595,6495)
\path(18810,6495)(17550,6495)
\path(18810,7395)(17550,7395)
\path(18765,5280)(17505,5280)
\path(17527,5280)(18810,6495)
\path(17527,5303)(18810,7395)
\path(18765,5280)(17550,3390)
\path(18810,3390)(17595,3390)
\path(18810,2490)(17550,2490)
\path(17550,2490)(18810,3390)
\path(17550,2490)(18765,5280)
\thicklines
\path(13635,1500)(13634,1501)(13632,1504)
	(13627,1509)(13620,1517)(13610,1528)
	(13597,1543)(13581,1562)(13561,1586)
	(13537,1613)(13510,1645)(13480,1680)
	(13447,1719)(13411,1762)(13374,1806)
	(13335,1853)(13296,1901)(13255,1951)
	(13215,2001)(13175,2051)(13135,2101)
	(13096,2151)(13059,2200)(13022,2248)
	(12987,2295)(12954,2342)(12922,2387)
	(12891,2431)(12862,2474)(12834,2516)
	(12808,2557)(12783,2598)(12759,2638)
	(12737,2677)(12716,2717)(12696,2755)
	(12677,2794)(12658,2833)(12641,2873)
	(12624,2912)(12608,2952)(12593,2992)
	(12579,3030)(12566,3068)(12553,3107)
	(12540,3146)(12528,3187)(12517,3228)
	(12505,3271)(12494,3315)(12484,3360)
	(12473,3407)(12463,3456)(12453,3507)
	(12443,3560)(12433,3615)(12423,3672)
	(12414,3731)(12404,3793)(12395,3857)
	(12385,3923)(12376,3991)(12367,4062)
	(12357,4134)(12348,4208)(12339,4283)
	(12330,4360)(12321,4436)(12313,4512)
	(12305,4587)(12297,4661)(12289,4733)
	(12282,4801)(12276,4866)(12269,4927)
	(12264,4983)(12259,5033)(12255,5078)
	(12251,5116)(12248,5149)(12245,5176)
	(12243,5197)(12242,5212)(12241,5223)
	(12240,5230)(12240,5233)(12240,5235)
\path(13590,1590)(13589,1591)(13588,1594)
	(13586,1600)(13582,1609)(13576,1622)
	(13568,1639)(13558,1661)(13546,1689)
	(13532,1721)(13515,1759)(13496,1801)
	(13475,1849)(13452,1901)(13427,1957)
	(13401,2016)(13374,2079)(13345,2144)
	(13316,2210)(13287,2278)(13257,2347)
	(13227,2416)(13198,2485)(13169,2552)
	(13140,2619)(13112,2685)(13086,2749)
	(13060,2812)(13034,2873)(13010,2932)
	(12987,2989)(12965,3044)(12944,3097)
	(12924,3148)(12905,3198)(12887,3246)
	(12870,3293)(12853,3338)(12838,3381)
	(12823,3424)(12809,3465)(12795,3506)
	(12783,3546)(12771,3584)(12759,3623)
	(12748,3661)(12738,3698)(12728,3735)
	(12716,3779)(12705,3823)(12694,3867)
	(12684,3911)(12675,3955)(12665,3999)
	(12657,4043)(12648,4087)(12640,4131)
	(12632,4176)(12625,4220)(12618,4265)
	(12611,4310)(12605,4354)(12599,4399)
	(12593,4444)(12588,4489)(12583,4533)
	(12578,4577)(12574,4621)(12569,4664)
	(12565,4707)(12562,4749)(12558,4791)
	(12555,4832)(12552,4873)(12549,4912)
	(12546,4951)(12543,4989)(12541,5026)
	(12538,5062)(12536,5098)(12534,5132)
	(12532,5166)(12529,5199)(12527,5232)
	(12525,5263)(12522,5294)(12520,5325)
	(12518,5355)(12514,5392)(12511,5429)
	(12507,5465)(12503,5501)(12498,5538)
	(12494,5575)(12488,5612)(12483,5650)
	(12477,5689)(12470,5729)(12463,5771)
	(12456,5814)(12448,5859)(12440,5905)
	(12431,5952)(12422,6000)(12412,6049)
	(12402,6099)(12393,6147)(12383,6194)
	(12374,6239)(12365,6281)(12357,6320)
	(12350,6353)(12344,6382)(12340,6405)
	(12336,6423)(12333,6436)(12331,6444)
	(12330,6448)(12330,6450)
\path(13635,1545)(13634,1546)(13633,1550)
	(13631,1556)(13627,1565)(13622,1579)
	(13614,1598)(13605,1621)(13594,1650)
	(13580,1684)(13565,1723)(13548,1767)
	(13529,1816)(13509,1868)(13488,1925)
	(13465,1984)(13442,2045)(13418,2109)
	(13394,2173)(13370,2238)(13346,2303)
	(13322,2367)(13299,2431)(13277,2494)
	(13255,2555)(13234,2615)(13214,2674)
	(13195,2731)(13176,2786)(13159,2840)
	(13142,2893)(13127,2944)(13112,2993)
	(13098,3042)(13084,3089)(13071,3136)
	(13059,3182)(13048,3227)(13037,3272)
	(13027,3316)(13017,3361)(13008,3405)
	(12999,3450)(12990,3495)(12982,3538)
	(12974,3582)(12967,3626)(12960,3670)
	(12953,3715)(12946,3761)(12940,3807)
	(12933,3854)(12927,3901)(12921,3949)
	(12915,3998)(12909,4047)(12904,4097)
	(12898,4148)(12893,4199)(12887,4250)
	(12882,4302)(12877,4355)(12872,4407)
	(12867,4460)(12862,4513)(12857,4566)
	(12853,4619)(12848,4672)(12843,4725)
	(12839,4777)(12834,4830)(12830,4882)
	(12825,4933)(12821,4984)(12816,5034)
	(12812,5083)(12807,5132)(12803,5181)
	(12798,5228)(12793,5275)(12789,5320)
	(12784,5365)(12779,5410)(12774,5453)
	(12769,5496)(12764,5538)(12759,5579)
	(12754,5620)(12748,5660)(12743,5700)
	(12736,5745)(12728,5790)(12721,5834)
	(12713,5878)(12705,5922)(12696,5966)
	(12687,6010)(12677,6055)(12666,6100)
	(12655,6146)(12644,6194)(12631,6242)
	(12618,6291)(12604,6342)(12590,6394)
	(12574,6448)(12558,6503)(12541,6559)
	(12524,6616)(12506,6674)(12488,6733)
	(12469,6792)(12450,6851)(12432,6909)
	(12413,6966)(12395,7020)(12378,7072)
	(12362,7120)(12348,7165)(12334,7205)
	(12323,7240)(12312,7270)(12304,7294)
	(12297,7314)(12292,7328)(12289,7339)
	(12287,7345)(12286,7348)(12285,7350)
\path(13635,2445)(13634,2447)(13633,2450)
	(13631,2457)(13627,2468)(13621,2483)
	(13614,2504)(13604,2530)(13592,2563)
	(13579,2600)(13563,2644)(13545,2693)
	(13525,2747)(13504,2805)(13482,2866)
	(13458,2931)(13433,2998)(13408,3066)
	(13382,3136)(13357,3205)(13331,3274)
	(13305,3342)(13280,3409)(13256,3474)
	(13232,3538)(13209,3599)(13186,3658)
	(13164,3714)(13143,3769)(13123,3820)
	(13103,3870)(13084,3917)(13066,3962)
	(13048,4005)(13031,4046)(13014,4086)
	(12998,4123)(12982,4159)(12967,4194)
	(12951,4227)(12937,4259)(12922,4290)
	(12907,4321)(12893,4350)(12873,4389)
	(12853,4427)(12833,4464)(12813,4500)
	(12792,4535)(12772,4569)(12750,4603)
	(12728,4637)(12705,4671)(12682,4705)
	(12657,4739)(12631,4774)(12604,4809)
	(12577,4844)(12548,4880)(12518,4917)
	(12488,4953)(12458,4988)(12428,5023)
	(12399,5057)(12371,5089)(12345,5119)
	(12321,5145)(12300,5169)(12283,5189)
	(12268,5204)(12257,5217)(12249,5225)
	(12244,5231)(12241,5234)(12240,5235)
\path(13635,2445)(13635,2446)(13634,2450)
	(13632,2456)(13630,2466)(13627,2480)
	(13623,2499)(13617,2523)(13610,2553)
	(13602,2589)(13592,2630)(13581,2678)
	(13568,2731)(13555,2789)(13540,2852)
	(13524,2919)(13507,2990)(13490,3064)
	(13472,3140)(13453,3219)(13434,3298)
	(13415,3378)(13396,3458)(13377,3537)
	(13358,3616)(13339,3693)(13321,3769)
	(13302,3843)(13285,3915)(13267,3984)
	(13250,4052)(13234,4117)(13218,4180)
	(13202,4241)(13187,4299)(13172,4356)
	(13157,4410)(13143,4462)(13130,4512)
	(13116,4561)(13103,4607)(13090,4653)
	(13077,4696)(13064,4739)(13052,4780)
	(13039,4820)(13027,4859)(13015,4898)
	(13002,4935)(12990,4972)(12975,5018)
	(12959,5063)(12943,5106)(12927,5150)
	(12911,5193)(12895,5235)(12878,5277)
	(12860,5320)(12842,5362)(12824,5405)
	(12805,5448)(12785,5492)(12764,5537)
	(12742,5583)(12720,5629)(12697,5677)
	(12673,5725)(12648,5775)(12622,5825)
	(12596,5875)(12570,5926)(12543,5977)
	(12516,6028)(12490,6077)(12464,6125)
	(12439,6172)(12415,6216)(12392,6257)
	(12372,6294)(12353,6328)(12337,6357)
	(12323,6383)(12311,6403)(12302,6420)
	(12295,6432)(12290,6440)(12287,6446)
	(12286,6449)(12285,6450)
\path(13635,2445)(13635,2446)(13634,2449)
	(13634,2455)(13632,2464)(13631,2478)
	(13628,2496)(13625,2519)(13621,2548)
	(13617,2583)(13611,2623)(13605,2670)
	(13598,2722)(13590,2780)(13582,2843)
	(13572,2911)(13563,2984)(13552,3060)
	(13541,3140)(13530,3223)(13518,3307)
	(13506,3393)(13493,3481)(13481,3568)
	(13468,3656)(13456,3743)(13443,3829)
	(13430,3914)(13418,3997)(13406,4079)
	(13393,4158)(13381,4236)(13370,4311)
	(13358,4384)(13346,4455)(13335,4523)
	(13324,4590)(13313,4654)(13302,4716)
	(13291,4775)(13280,4833)(13270,4889)
	(13259,4943)(13248,4995)(13238,5046)
	(13227,5096)(13217,5144)(13206,5190)
	(13195,5236)(13185,5281)(13174,5325)
	(13163,5368)(13151,5411)(13140,5452)
	(13126,5503)(13111,5553)(13097,5603)
	(13081,5651)(13066,5700)(13049,5748)
	(13033,5796)(13015,5844)(12997,5893)
	(12979,5942)(12959,5991)(12938,6041)
	(12917,6092)(12895,6144)(12871,6196)
	(12847,6250)(12822,6305)(12795,6362)
	(12768,6419)(12740,6477)(12711,6536)
	(12681,6596)(12651,6656)(12620,6717)
	(12590,6777)(12559,6836)(12529,6895)
	(12499,6951)(12471,7005)(12443,7057)
	(12418,7105)(12394,7149)(12373,7189)
	(12353,7225)(12337,7255)(12323,7281)
	(12311,7303)(12302,7319)(12295,7332)
	(12290,7340)(12287,7346)(12286,7349)(12285,7350)
\thinlines
\path(12240,5235)(13680,6450)
\path(12240,5235)(13590,7305)
\path(12240,5235)(13680,5685)
\path(13635,7350)(12285,6450)
\path(13680,6450)(12285,6450)
\path(12240,5235)(10980,5235)
\path(11002,5235)(12285,6450)
\path(11002,5258)(12285,7350)
\path(4297,5258)(5580,7350)
\path(4297,5235)(5580,6450)
\path(5535,5235)(4275,5235)
\path(6975,6450)(5580,6450)
\path(7020,1545)(8235,4335)
\path(7020,1545)(8280,2445)
\path(8280,1545)(7020,1545)
\path(8280,2445)(7020,2445)
\path(8235,4335)(7065,2445)
\path(7020,4335)(8235,4335)
\path(5580,7350)(6975,7350)
\path(5580,7350)(6975,6450)
\path(5580,7350)(6975,5685)
\path(5580,6450)(6975,5685)
\path(14895,4335)(13680,4335)
\path(13680,4335)(14850,2445)
\path(13635,2445)(14895,2445)
\path(13635,1545)(14895,1545)
\path(14895,1545)(13635,2445)
\path(14895,1545)(13680,4335)
\thicklines
\path(12285,7350)(12286,7349)(12289,7346)
	(12294,7340)(12302,7331)(12313,7318)
	(12328,7301)(12346,7280)(12368,7254)
	(12394,7224)(12423,7191)(12455,7153)
	(12489,7113)(12525,7070)(12563,7025)
	(12601,6979)(12640,6932)(12679,6885)
	(12717,6838)(12754,6791)(12790,6745)
	(12825,6701)(12859,6657)(12891,6615)
	(12921,6573)(12950,6533)(12978,6494)
	(13004,6456)(13029,6418)(13052,6382)
	(13075,6346)(13096,6310)(13117,6275)
	(13136,6240)(13155,6205)(13173,6169)
	(13190,6134)(13208,6097)(13223,6064)
	(13237,6031)(13252,5997)(13266,5962)
	(13280,5926)(13293,5889)(13307,5851)
	(13320,5812)(13334,5772)(13347,5729)
	(13361,5685)(13374,5640)(13388,5592)
	(13402,5542)(13416,5490)(13430,5436)
	(13444,5380)(13459,5322)(13474,5262)
	(13489,5200)(13504,5137)(13519,5072)
	(13534,5007)(13549,4942)(13563,4877)
	(13578,4813)(13591,4751)(13604,4691)
	(13617,4634)(13628,4582)(13638,4534)
	(13647,4491)(13655,4453)(13662,4421)
	(13668,4395)(13672,4374)(13675,4358)
	(13677,4347)(13679,4340)(13680,4337)(13680,4335)
\path(13680,4290)(13679,4292)(13678,4296)
	(13675,4303)(13670,4315)(13663,4332)
	(13654,4354)(13642,4382)(13628,4416)
	(13612,4455)(13593,4500)(13572,4550)
	(13550,4603)(13526,4661)(13501,4721)
	(13475,4782)(13448,4845)(13421,4908)
	(13394,4971)(13368,5033)(13341,5094)
	(13316,5153)(13291,5209)(13266,5264)
	(13243,5316)(13220,5366)(13198,5413)
	(13177,5458)(13157,5501)(13138,5541)
	(13119,5580)(13101,5616)(13083,5650)
	(13066,5683)(13049,5714)(13032,5743)
	(13016,5772)(13000,5799)(12984,5825)
	(12968,5850)(12946,5882)(12925,5913)
	(12903,5943)(12881,5971)(12859,5999)
	(12836,6025)(12812,6051)(12788,6077)
	(12762,6102)(12735,6127)(12706,6152)
	(12677,6177)(12646,6202)(12614,6227)
	(12580,6252)(12547,6276)(12513,6301)
	(12479,6324)(12446,6346)(12415,6367)
	(12386,6386)(12361,6402)(12339,6417)
	(12321,6428)(12307,6437)(12297,6443)
	(12290,6447)(12287,6449)(12285,6450)
\path(13680,4290)(13678,4292)(13675,4295)
	(13668,4301)(13657,4311)(13643,4325)
	(13623,4342)(13600,4364)(13572,4390)
	(13541,4418)(13507,4450)(13470,4483)
	(13432,4518)(13393,4553)(13354,4589)
	(13315,4624)(13277,4657)(13241,4690)
	(13205,4721)(13172,4750)(13140,4778)
	(13109,4804)(13080,4828)(13053,4851)
	(13027,4872)(13002,4892)(12978,4910)
	(12956,4928)(12933,4944)(12912,4959)
	(12891,4974)(12870,4987)(12845,5004)
	(12820,5019)(12795,5034)(12770,5047)
	(12744,5061)(12718,5073)(12691,5086)
	(12663,5098)(12634,5110)(12603,5122)
	(12571,5133)(12537,5145)(12503,5157)
	(12468,5168)(12433,5179)(12398,5190)
	(12365,5200)(12335,5208)(12309,5216)
	(12286,5222)(12268,5227)(12255,5231)
	(12247,5233)(12242,5234)(12240,5235)
\put(6480,1455){\makebox(0,0)[b]{\smash{{{\SetFigFont{5}{6.0}{rm}$b_{\frac{n}{2}}$}}}}}
\put(0,2355){\makebox(0,0)[b]{\smash{{{\SetFigFont{5}{6.0}{rm}$b_{\frac{n}{2}}$}}}}}
\put(18270,105){\makebox(0,0)[b]{\smash{{{\SetFigFont{5}{6.0}{rm}$G*b_i$}}}}}
\put(14220,105){\makebox(0,0)[b]{\smash{{{\SetFigFont{5}{6.0}{rm}$G*b_i$}}}}}
\put(7560,195){\makebox(0,0)[b]{\smash{{{\SetFigFont{5}{6.0}{rm}$G$}}}}}
\put(19260,2445){\makebox(0,0)[b]{\smash{{{\SetFigFont{5}{6.0}{rm}$b_{i+1}$}}}}}
\put(19350,3435){\makebox(0,0)[b]{\smash{{{\SetFigFont{5}{6.0}{rm}$b_{\frac{n}{2}-1}$}}}}}
\put(19170,4065){\makebox(0,0)[b]{\smash{{{\SetFigFont{5}{6.0}{rm}$b_{\frac{n}{2}}$}}}}}
\put(17145,2535){\makebox(0,0)[b]{\smash{{{\SetFigFont{5}{6.0}{rm}$w_{\frac{n}{2}-2}$}}}}}
\put(17010,6405){\makebox(0,0)[b]{\smash{{{\SetFigFont{5}{6.0}{rm}$b_{i-2}$}}}}}
\put(17010,7305){\makebox(0,0)[b]{\smash{{{\SetFigFont{5}{6.0}{rm}$b_{i-1}$}}}}}
\put(17010,5685){\makebox(0,0)[b]{\smash{{{\SetFigFont{5}{6.0}{rm}$b_1$}}}}}
\put(19440,6405){\makebox(0,0)[b]{\smash{{{\SetFigFont{5}{6.0}{rm}$w_{\frac{n}{2}-i}$}}}}}
\put(0,3255){\makebox(0,0)[b]{\smash{{{\SetFigFont{5}{6.0}{rm}$b_{\frac{n}{2}-1}$}}}}}
\put(0,7260){\makebox(0,0)[b]{\smash{{{\SetFigFont{5}{6.0}{rm}$b_1$}}}}}
\put(0,6360){\makebox(0,0)[b]{\smash{{{\SetFigFont{5}{6.0}{rm}$b_2$}}}}}
\put(0,5235){\makebox(0,0)[b]{\smash{{{\SetFigFont{5}{6.0}{rm}$b_i$}}}}}
\put(2430,7260){\makebox(0,0)[b]{\smash{{{\SetFigFont{5}{6.0}{rm}$w_{\frac{n}{2}}$}}}}}
\put(2430,6360){\makebox(0,0)[b]{\smash{{{\SetFigFont{5}{6.0}{rm}$w_{\frac{n}{2}-1}$}}}}}
\put(2295,5280){\makebox(0,0)[b]{\smash{{{\SetFigFont{5}{6.0}{rm}$w_{\frac{n}{2}-i+1}$}}}}}
\put(2430,2355){\makebox(0,0)[b]{\smash{{{\SetFigFont{5}{6.0}{rm}$w_1$}}}}}
\put(2430,3255){\makebox(0,0)[b]{\smash{{{\SetFigFont{5}{6.0}{rm}$w_2$}}}}}
\put(17010,5280){\makebox(0,0)[b]{\smash{{{\SetFigFont{5}{6.0}{rm}$b_i$}}}}}
\put(19260,5325){\makebox(0,0)[b]{\smash{{{\SetFigFont{5}{6.0}{rm}$w_{\frac{n}{2}}$}}}}}
\put(19395,7485){\makebox(0,0)[b]{\smash{{{\SetFigFont{5}{6.0}{rm}$w_{\frac{n}{2}-i+1}$}}}}}
\put(17145,4020){\makebox(0,0)[b]{\smash{{{\SetFigFont{5}{6.0}{rm}$w_1$}}}}}
\put(17145,3345){\makebox(0,0)[b]{\smash{{{\SetFigFont{5}{6.0}{rm}$w_2$}}}}}
\put(990,105){\makebox(0,0)[b]{\smash{{{\SetFigFont{5}{6.0}{rm}$G$}}}}}
\put(13995,5685){\makebox(0,0)[b]{\smash{{{\SetFigFont{5}{6.0}{rm}$b_{i-1}$}}}}}
\put(13995,7305){\makebox(0,0)[b]{\smash{{{\SetFigFont{5}{6.0}{rm}$b_1$}}}}}
\put(14040,6405){\makebox(0,0)[b]{\smash{{{\SetFigFont{5}{6.0}{rm}$b_2$}}}}}
\put(10485,5235){\makebox(0,0)[b]{\smash{{{\SetFigFont{5}{6.0}{rm}$b_i$}}}}}
\put(12285,7800){\makebox(0,0)[b]{\smash{{{\SetFigFont{5}{6.0}{rm}$w_{\frac{n}{2}}$}}}}}
\put(12375,6855){\makebox(0,0)[b]{\smash{{{\SetFigFont{5}{6.0}{rm}$w_{\frac{n}{2}-1}$}}}}}
\put(11745,4875){\makebox(0,0)[b]{\smash{{{\SetFigFont{5}{6.0}{rm}$w_{\frac{n}{2}-i+1}$}}}}}
\put(5580,4785){\makebox(0,0)[b]{\smash{{{\SetFigFont{5}{6.0}{rm}$w_{\frac{n}{2}-i+1}$}}}}}
\put(5670,6855){\makebox(0,0)[b]{\smash{{{\SetFigFont{5}{6.0}{rm}$w_{\frac{n}{2}-1}$}}}}}
\put(5580,7800){\makebox(0,0)[b]{\smash{{{\SetFigFont{5}{6.0}{rm}$w_{\frac{n}{2}}$}}}}}
\put(3780,5235){\makebox(0,0)[b]{\smash{{{\SetFigFont{5}{6.0}{rm}$b_i$}}}}}
\put(7335,6405){\makebox(0,0)[b]{\smash{{{\SetFigFont{5}{6.0}{rm}$b_2$}}}}}
\put(7290,7305){\makebox(0,0)[b]{\smash{{{\SetFigFont{5}{6.0}{rm}$b_1$}}}}}
\put(7290,5685){\makebox(0,0)[b]{\smash{{{\SetFigFont{5}{6.0}{rm}$b_{i-1}$}}}}}
\put(8910,2355){\makebox(0,0)[b]{\smash{{{\SetFigFont{5}{6.0}{rm}$w_2$}}}}}
\put(8910,1455){\makebox(0,0)[b]{\smash{{{\SetFigFont{5}{6.0}{rm}$w_1$}}}}}
\put(6570,4335){\makebox(0,0)[b]{\smash{{{\SetFigFont{5}{6.0}{rm}$b_{i+1}$}}}}}
\put(8685,4290){\makebox(0,0)[b]{\smash{{{\SetFigFont{5}{6.0}{rm}$w_{\frac{n}{2}-i}$}}}}}
\put(6480,2355){\makebox(0,0)[b]{\smash{{{\SetFigFont{5}{6.0}{rm}$b_{\frac{n}{2}-1}$}}}}}
\put(14400,4695){\makebox(0,0)[b]{\smash{{{\SetFigFont{5}{6.0}{rm}$w_{\frac{n}{2}-i}$}}}}}
\put(15345,4335){\makebox(0,0)[b]{\smash{{{\SetFigFont{5}{6.0}{rm}$b_{i+1}$}}}}}
\put(13005,1455){\makebox(0,0)[b]{\smash{{{\SetFigFont{5}{6.0}{rm}$w_1$}}}}}
\put(15435,2355){\makebox(0,0)[b]{\smash{{{\SetFigFont{5}{6.0}{rm}$b_4$}}}}}
\put(15435,1455){\makebox(0,0)[b]{\smash{{{\SetFigFont{5}{6.0}{rm}$b_5$}}}}}
\thinlines
\put(18810,2490){\blacken\ellipse{180}{180}}
\put(18810,2490){\ellipse{180}{180}}
\end{picture}
}